\documentclass[12pt]{article}
\usepackage{fullpage}
\usepackage[authoryear]{natbib}
\usepackage[utf8]{inputenc}
\usepackage{amsmath}
\usepackage{amssymb}
\usepackage{amsfonts,}
\usepackage{bbm}
\usepackage{graphicx}
\usepackage{float}
\usepackage{caption}
\usepackage{subcaption,mathrsfs,multirow}
\usepackage[ruled]{algorithm2e}
\usepackage{array}
\usepackage{bm}

\usepackage{verbatim}
\usepackage{listings}
\usepackage{color}
\newtheorem{theorem}{Theorem}[section]
\newtheorem{lemma}[theorem]{Lemma}
\usepackage{epsfig}
\usepackage{wasysym}

\newcommand{\bfZ}{\mathbf{Z}}

\makeatletter
\newcommand*\rel@kern[1]{\kern#1\dimexpr\macc@kerna}
\newcommand*\widebar[1]{%
	\begingroup
	\def\mathaccent##1##2{%
		\rel@kern{0.8}%
		\overline{\rel@kern{-0.8}\macc@nucleus\rel@kern{0.2}}%
		\rel@kern{-0.2}%
	}%
	\macc@depth\@ne
	\let\math@bgroup\@empty \let\math@egroup\macc@set@skewchar
	\mathsurround\z@ \frozen@everymath{\mathgroup\macc@group\relax}%
	\macc@set@skewchar\relax
	\let\mathaccentV\macc@nested@a
	\macc@nested@a\relax111{#1}%
	\endgroup
}
\makeatother
\newcommand*\diff{\mathop{}\!\mathrm{d}}
\usepackage{color}

\newcommand{\comout}[1]{}

\def\lpbox#1{\vskip1mm \begin{center}
        \hspace{.0\textwidth}\vbox{\hrule\hbox{\vrule\kern6pt
\parbox{.95\textwidth}{\kern6pt \blue #1 (LP)\vskip6pt}\kern6pt\vrule}\hrule}
        \end{center} \vskip-5mm}

\usepackage[usenames,dvipsnames,svgnames,table]{xcolor}
\usepackage[colorlinks=true,
            linkcolor=red,
            urlcolor=blue,
            citecolor=blue]{hyperref}

\usepackage{setspace}

\definecolor{lbcolor}{rgb}{0.95,0.95,0.95}
\definecolor{darkred}{RGB}{150,50,50}
\definecolor{brown}{RGB}{250,100,100}
\definecolor{green}{RGB}{000,150,100}
\definecolor{purple}{RGB}{200,000,250}

\def\blue{\color{blue}}

\def\trans{^{\scriptscriptstyle \sf T}}
\def\Beta{\bm \beta}
\def\Phibar{\bm{\bar{\Phi}}}
\def\Shat{\widehat{S}}
\def\Sighat{\widehat{\Sigma}}
\def\D{\mbox{\tiny D}}
\def\ID{\mbox{\tiny ID}}

\def\k{\textit{-k}}
\def\L{\mbox{\tiny L}}
\def\IL{\mbox{\tiny IL}}
\def\U{\mbox{\tiny U}}
\def\IU{\mbox{\tiny IU}}
\def\ssd{{\mbox{\tiny SSD}}(t)}
\def\ssl{{\mbox{\tiny SSL}}(t)}
\def\ssu{{\mbox{\tiny SSU}}(t)}
\def\ssc{{\mbox{\tiny SSC}}(t)}
\def\sd{{\mbox{\tiny SD}}(t)}
\def\sl{{\mbox{\tiny SL}}(t)}
\def\su{{\mbox{\tiny SU}}(t)}
\def\sc{{\mbox{\tiny SC}}(t)}
\def\NL{{\textit{\tiny NL}}}
\def\NU{{\textit{\tiny NU}}}
\def\N{{\textit{\tiny N}}}
\def\dob{\bm\cdot}
\def\S{{\mbox{\tiny S}}(t)}
\def\SS{{\mbox{\tiny SS}}(t)}

\def\cv{{\mbox{\tiny cv}}}
\def\Mhat{\widehat{\mathbf{M}}}
\def\Vhat{\widehat{\mbox{V}}}
\def\mhat{\hat{\mbox{m}}}

\begin{document}

\begin{center}
{\large \bf Semi-supervised Estimation of Event Rate with Doubly-censored Survival Data} \vspace{.2in}
\end{center}

\begin{center}\vspace{-2mm}
{Yang Wang \footnote{Co-first author}} \vspace{-2mm}
\end{center}
{Department of Biomedical Informatics, Harvard University, Boston, Massachusetts 02115 U.S.A. }

\begin{center}\vspace{-2mm}
{Qingning Zhou \footnote{Co-first author}}\vspace{-2mm}
\end{center}
{Department of Mathematics and Statistics, the University of North Carolina at Charlotte, Charlotte, NC 28223 U.S.A.}

\begin{center}\vspace{-2mm}
{Tianxi Cai }\vspace{-2mm}
\end{center}
{Department of Biostatistics $\&$ Biomedical Informatics, Harvard University, Boston, Massachusetts 02115 U.S.A. }

\begin{center}\vspace{-2mm}
{Xuan Wang \footnote{Corresponding author: xwang@hsph.harvard.edu}}\vspace{-2mm}
\end{center}
{Department of Biostatistics $\&$ Biomedical Informatics, Harvard University, Boston, Massachusetts 02115 U.S.A.}

\ \

\begin{abstract}

Electronic Health Record (EHR) has emerged as a valuable source of data for translational research. To leverage EHR data for risk prediction and subsequently clinical decision support, clinical endpoints are often time to onset of a clinical condition of interest. Precise information on clinical event times are often not directly available and requires labor-intensive manual chart review to ascertain. In addition, events may occur outside of the hospital system, resulting in both left and right censoring or often termed double censoring. On the other hand, proxies such as time to the first diagnostic code are readily available yet with varying degrees of accuracy. Using error-prone event times derived from these proxies can lead to biased risk estimates while only relying on manually annotated event times, which are typically only available for a small subset of patients, can lead to high variability. This signifies the need for semi-supervised estimation methods that can efficiently combine information from both the small subset of labeled observations and a large size of surrogate proxies. While semi-supervised estimation methods have been recently developed for binary and right-censored data, no methods currently exist in the presence of double censoring. This paper fills the gap by developing a robust and efficient  {\bf S}emi-supervised {\bf E}stimation of {\bf E}vent rate with {\bf D}oubly-censored {\bf S}urvival data (SEEDS) by leveraging a small set of gold standard labels and a large set of surrogate features. Under mild regularity conditions, we demonstrate that the proposed SEEDS estimator is consistent and asymptotically normal. Extensive simulation results illustrate that SEEDS performs well in finite samples and can be substantially more efficient compared to the supervised counterpart. We apply the SEEDS procedure to estimate the age-specific survival rate of type 2 diabetes (T2D) using EHR data from Mass General Brigham (MGB). 
\end{abstract}

\newpage
\clearpage

\setstretch{1.1}

\section{Introduction}

Electronic Health Record (EHR) has emerged as an increasing and valuable source of data for translational research \citep{Hripcsak2013ehr, Miotto2016ehr, Hodgkins2018ehr}. It contains longitudinal information on diagnoses, procedures, billing codes, laboratory tests, prescriptions, and free-text clinical notes for a large number of patients. Accurate personalized EHR-based risk prediction models have high translational values since they are built with EHR entities and can readily be integrated into the health system to improve clinical decision-making. However, risk assessment with EHR data remains challenging. Clinical events may occur outside the hospital system, resulting in left- or right-censored event times, known as doubly-censored data. Another challenge is that onset time information of clinical events is often embedded in free-text clinical notes and requires labor-intensive manual chart review to ascertain. Relying only on a small set of precise event times from manual chart review can lead to high variability. On the other hand, error-prone event times derived from proxies such as time to the first diagnostic code, while readily available for a large number of patients, may lead to biased results. It is thus desirable to effectively combine information from a small set of gold-standard event times and a large set of proxies. 

For example, EHR data on Type 2 Diabetes (T2D) from the Massachusetts General Brigham (MGB) health system include 115,236 patients who have at least one International Classification of Diseases (ICD) code. The precise onset times of T2D are ascertained by manual chart review for 1,613 patients, of which 44 patients have already developed T2D at the first visit, resulting in left-censored observations, and 1401 patients have not yet developed T2D at the last visit, yielding right-censored observations. Proxies such as time to the first T2D code (PheCode:250.2) are available for all 115,236 patients.

The analysis of doubly-censored survival data has been investigated by many authors. Among others, \citet{Turnbull1974}, \citet{Tsai1985}, \citet{Gu1993}, and \citet{Messaci2011} have considered non-parametric estimation of survival rate based on doubly-censored data, while \citet{CaiCheng2004}, \citet{Sun2004Reg}, \citet{Zhang2009Imp}, and \citet{Shuwei2020reg} have studied semi-parametric regression analysis of doubly-censored data. All these methods require precise event times and cannot leverage the information on proxies or surrogates available on a large number of patients in EHR. 

Semi-supervised learning has recently attracted a great deal of attention and has also been applied to EHR research. It can efficiently combine information from a small set of labeled data with precise outcomes and a large set of unlabeled data with only proxies or surrogate outcomes. The existing semi-supervised methods mainly focus on binary, continuous, or right-censored survival outcomes. For example,  \citet{Garla2013semi}, \citet{Dligach2015semi}, \citet{Beaulieu2016semi}, \citet{Henderson2018semi}, and \citet{Yichi2019semi} developed semi-supervised algorithms for classification based on machine learning approaches to study EHR phenotyping, while \citet{wang2012extracting}, \citet{Perez2017semi}, \citet{Sanchez2022semi}, and \citet{WangNi2021semi} considered semi-supervised classification in other settings. \citet{Gronsbell2018semi} and \citet{gronsbell2022efficient} developed semi-supervised approaches for evaluating model prediction performance with binary outcomes. \citet{Chakrabortty2018effic}, \citet{Anru2019semi}, and \citet{Cheng2021robust} proposed semi-supervised inference procedures for regression analysis with continuous outcomes. \citet{Chai2017semi} developed a semi-supervised method under the Cox model to predict survival rate with right-censored data. \citet{Ahuja2021semi} considered semi-supervised estimation for survival rate using the IPCW approach based on current status data. To the best of our knowledge, no semi-supervised methods are available in the presence of double censoring.

In this paper, to estimate the event rate based on doubly-censored EHR data by leveraging a small set of precise event times and a large set of proxies or surrogate features, we propose a robust and efficient method for {\bf S}emi-supervised {\bf E}stimation of {\bf E}vent rate with {\bf D}oubly-censored {\bf S}urvival data (SEEDS). The proposed estimation procedure has two steps. In Step 1, we first construct a semi-supervised estimator of the survival rate using doubly-censored EHR data. To fully utilize the data, we create three specific types of labels from the doubly-censored labels and then derive three corresponding semi-supervised estimators. In Step 2, we construct the SEEDS estimator as an optimal linear combination of the three semi-supervised estimators obtained in Step 1. We also present its supervised counterpart that only uses the labeled data as a benchmark. We show that the proposed SEEDS estimator is consistent and asymptotically normal under mild regularity conditions, and also demonstrate that it is more efficient than its supervised counterpart. In addition, since we repeatedly use the labeled data in calculating our SEEDS estimator, we adopt a cross-fitting approach to correct for the potential over-fitting bias in small samples.

The remainder of this paper is organized as follows. In Section~\ref{method}, we develop three semi-supervised estimators based on three types of labels, respectively, and then construct the SEEDS estimator by combining the three semi-supervised estimators. We also establish the asymptotic properties of the proposed SEEDS estimator and present a cross-fitting approach to correct for the potential over-fitting bias in small samples. Extensive simulation studies are performed in Section~\ref{simul} and an application to EHR data on T2D is provided in Section~\ref{applic}. Some discussions are included in Section~\ref{dissc}. Proofs of the asymptotic properties and more simulation results are in the Appendix.

\section{Proposed estimation method}
\label{method}
Suppose that there are $n+N$ individuals in the EHR cohort. For individual $i$, let $T_i$ denote the event time of interest subject to double censoring. Let $L_i$ and $U_i$ denote the left and right censoring times, respectively, that are always observed and satisfy $P(L_i < U_i)=1$. Under double censoring, $T_i$ is observed only if it falls within the interval $\left[L_i, U_i\right]$, while $T_i$ is left-censored if $T_i < L_i$ and right-censored if $T_i > U_i$. Thus, the observed data for $T_i$ can be represented by $\big\{X_i=\max(L_i,\min(T_i,U_i)),\delta_i\big\}$, where 
\begin{equation*}
    \delta_i=\begin{cases}
      1,   & \text{if} \quad L_i\leqslant  T_i\leqslant  U_i,\\
      2,   & \text{if} \quad T_i>U_i,\\
      3,   & \text{if} \quad T_i<L_i.
    \end{cases}
   \end{equation*}
Suppose there is a vector of $q$ surrogate outcomes $\mathbf{T}_{i}^{\ast}=(T_{1i}^{\ast},\ldots,T_{qi}^{\ast})^{\trans}$, which are proxies of $T_i$ and also subject to double censoring. Similarly, for $\mathbf{T}_{i}^{\ast}$, we only observe $\mathbf{X}_{i}^{\ast}=(X_{1i}^{\ast},\ldots,X_{qi}^{\ast})^{\trans}$ and $\bm{\delta}_{i}^{\ast}=(\delta_{1i}^{\ast},\ldots,\delta_{qi}^{\ast})^{\trans}$, where $X_{ki}^{\ast}=\max(L_i,\min(T_{ki}^{\ast},U_i))$ and 
   \begin{equation*}
    \delta_{ki}^{\ast}=\begin{cases}
      1,   & \text{if} \quad L_i\leqslant  T_{ki}^{\ast}\leqslant  U_i,\\
      2,   & \text{if} \quad T_{ki}^{\ast}>U_i,\\
      3,   & \text{if} \quad T_{ki}^{\ast}<L_i.
    \end{cases}
   \end{equation*}
for $k = 1, \ldots, q$. Suppose there is a $p_1$-dimensional vector of baseline covariate $\bfZ_i$.
Also, suppose there is a $p_2$-dimensional vector of covariate processes $\big\{\mathbf{z}_{i}(u): u \in [L_i, U_i]\big\}$. Let $\mathbf{Z}_{L_i}^t = \int_{L_i}^t \mathbf{z}_{i}(u)\, \diff V(u)$  denote the cumulative covariate processes from $L_i$ to $t$, where $V(u)$ is a stochastic process. Assume we observe the covariate processes up to $U_i$, that is, $\mathbf{\widebar{Z}}_{L_i}=\big\{\mathbf{Z}_{L_i}^t: L_i \leqslant t \leqslant U_i\big\}$. 
Moreover, we assume that $\big(T_i,\mathbf{T}_{i}^{\ast}, \bfZ_i, \mathbf{\widebar{Z}}_{L_i} \big)$ are independent of $\big(L_i, U_i\big)$. 
Suppose that the true outcome $\{X_i, \delta_i\}$ is observed only for a small set of labeled data of size $n$. Without loss of generality, we assume the labeled data consists of the first $n$ individuals in the cohort. Thus, we observe
$$
\mathcal{D}_{l}=\left\{\big(X_i,\delta_i,L_i,U_i,\mathbf{X}_{i}^{\ast\trans},\bm{\delta}_{i}^{\ast\trans}, \bfZ_i^{\trans}, \mathbf{\widebar{Z}}_{L_i}^{\trans}\big)^{\trans},\,\,i=1,\ldots,n\right\}
$$
for labeled data and 
$$
\mathcal{D}_{u}=\left\{\big(L_i,U_i,\mathbf{X}_{i}^{\ast\trans},\bm{\delta}_{i}^{\ast\trans},  \bfZ_i^{\trans}, \mathbf{\widebar{Z}}_{L_i}^{\trans}\big)^{\trans},\,\,i=n+1,\ldots,n+N\right\}
$$
for unlabeled data. Furthermore, we assume that $\log(N)/\log(n)\to \kappa_0 > 1$ as $n\to \infty$. Our goal is to develop a semi-supervised estimator of the survival function $S(t)={\rm Pr}(T\geqslant t)$ using both $\mathcal{D}_{l}$ and $\mathcal{D}_{u}$.

Since $L_i$ and $U_i$ are always observed, the left censoring label $\big\{L_i, \,I(T_i < L_i)\big\}$ and the right censoring label $\big\{U_i, \,I(T_i > U_i)\big\}$ can be treated as current status data in survival literature. In the following, we develop our estimation procedure in two steps. In Step 1, we construct three semi-supervised estimators of the survival function by using three types of labels, the exact observed event time label $\big(X_i, \delta_i\big)$, and the two current status labels, respectively. In particular, for each type of label, we first fit a working model that relates the event time to surrogate outcomes and covariates based on the labeled data $\mathcal{D}_l$, and then we obtain an estimator of the survival function based on the imputed risk of unlabeled data $\mathcal{D}_u$ from the fitted model. In Step 2, we construct the proposed SEEDS estimator as an optimal linear combination of the three estimators from Step 1, which is more efficient than each. Below, we describe the two steps of our estimation procedure in more detail. For comparison, we also present the supervised counterpart that only uses the labeled data $\mathcal{D}_l$. In addition, we develop a cross-fitting method to correct for the potential over-fitting bias with small samples.

\subsection{Step 1: Estimation based on three types of labels}

For each type of label, we develop a semi-supervised estimator of the survival function $S(t)$. Using labeled data $\mathcal{D}_l$, we firstly fit a working time-specific logistic model that relates the event time to some functions of surrogate outcomes $(\mathbf{X}_{i}^{\ast},\bm{\delta}_{i}^{\ast})$, baseline covariates $\bfZ_i$, and time-dependent covariates $\mathbf{Z}_{L_i}^t$. Then we obtain an estimator of the survival rate by marginalizing the imputed risk of unlabeled data $\mathcal{D}_u$ from the fitted model. We further improve the efficiency via intrinsic estimation \citep{Tan2010bounded}. More details are given below.

\subsubsection{Estimation based on the exact observed event time label}
\label{sect21}

In this section, we develop a semi-supervised estimator of the survival function $S(t)$ by using the exact observed event time label $\big(X_i, \delta_i\big)$. Recall that $T$ and $(L, U)$ are independent, then the survival function can be written as
\begin{equation*}
\begin{split}
S(t)&={\rm Pr}(T\geqslant t)\\
&={\rm Pr}(X\geqslant t> L\mid U\geqslant t>L)\\
&=E\Big[E\big[I(X\geqslant t>L)\mid \widebar{\textbf{W}}_{\D}^{t}, U\geqslant t>L\big]\mid U\geqslant t>L\Big],
\end{split}
\end{equation*}
where $\widebar{\textbf{W}}_{\D}^{t}=\big({1},\textbf{W}_{\D}^{t\trans}\big)^{\trans}$and $\textbf{W}_{\D}^{t}=\big(\mathbf{X}^{\ast\trans},\bm{\delta}^{\ast\trans}, \bfZ^{\trans},\mathbf{Z}_L^{t\trans}\big)^{\trans}$.

First, using labeled data $\mathcal{D}_l$, we fit a working time-specific logistic model as follows,
\begin{equation}
\label{dmodel}
E\left[I(X\geqslant t>L)\mid \widebar{\textbf{W}}_{\D}^{t}, U\geqslant t>L\right]=g(\Beta_{\D}^{t\trans}\Phibar_{\D}^{t}),
\end{equation}
where $\Beta_{\D}^{t}$ is an unknown regression parameter that depends on time $t$, $\Phibar_{\D}^{t} =\big(1,{{\bm\Phi}}(\textbf{W}_{\D}^{t})^{\trans}\big)^{\trans}$ with ${\bm\Phi}(\textbf{W}_{\D}^{t})$ being a basis function of $\textbf{W}_{\D}^{t}$, and $g(x)=\exp(x)/(1+\exp(x))$. Let $\widehat{\Beta}_{\D}^{t}$ denote the estimator of $\Beta_{\D}^{t}$ that solves the following estimating equation,
\begin{equation}
    \label{dsl}
    \mathbf{U}_n(\Beta_{\D}^{t})=\frac{1}{n}\sum_{i=1}^{n} \Phibar_{\D i}^{t} I(U_i\geqslant t>L_i)\big\{I(X_i\geqslant t>L_i)- g(\Beta_{\D}^{t\trans}\Phibar_{\D i}^{t})\big\}=0.
\end{equation}
Under some mild regularity conditions, we can show that $\widehat{\Beta}_{\D}^{t}$ is a consistent estimator for the unique solution, denoted by $\widebar{\Beta}_{\D}^t$, to $\widebar{\mathbf{U}}(\Beta_{\D}^{t})=E\big[\Phibar_{\D}^{t}I(U\geqslant t>L) \big\{ I(X\geqslant t>L)- g(\Beta_{\D}^{t\trans}\Phibar_{\D}^{t})\big\}\big]=0$. 

Next, we estimate the survival rate at time $t$ by marginalizing the imputed risk $g(\widehat{\Beta}_{\D}^{t\trans}\Phibar_{\D i}^{t})$ of unlabeled data $\mathcal{D}_u$ as follows, 
$$
\Shat_{\D}(t)=\frac{\sum_{i= n+1}^{n+N}I(U_i\geqslant t>L_i)g(\widehat{\Beta}_{\D}^{t \trans}\Phibar_{\D i}^{t})}{ \sum_{i=n+1}^{n+N}I(U_i\geqslant t>L_i)}.
$$
As shown in Appendix B, $\sqrt{n}\big(\Shat_{\D}(t)-S(t)\big)$ converges in distribution to a normal random variable with mean zero and variance 
\begin{equation}
\label{dsslasyvar}
\Sigma_{\D}(t)=I_{\D}(t)^{-2}E\bigl[I(U\geqslant t>L)\big\{I(X\geqslant t>L)-g(\widebar{\Beta}_{\D}^{t\trans}\Phibar_{\D}^{t})\big\}\bigr]^{2},
\end{equation}
where $I_{\D}(t)=E\left[I(U\geqslant t >L)\right]$. 

Lastly, we improve the efficiency of $\Shat_{\D}(t)$ using the idea of intrinsic efficiency in \citet{Tan2010bounded}. An alternative estimating equation for $\Beta_{\D}^t$ may be used to directly reduce the asymptotic variance of $\Shat_{\D}(t)$ when the imputation model \eqref{dmodel} is misspecified. For a fixed $\Phibar_{\D}^{t}$, we may find the estimating equation for $\Beta_{\D}^t$ that leads to the lowest asymptotic variance of the estimator of $S(t)$, which can be obtained by directly minimizing the variance \eqref{dsslasyvar} for $S(t)$. Specifically, we obtain the intrinsic estimator of $\Beta_{\D}^{t}$ as 
\begin{align}
\label{dintreq}
    \widehat{\Beta}_{\ID}^{t}=& \arg\min\limits_{\Beta_{\D}^{t}}\, \frac{1}{n}\sum_{i=1}^{n} \widebar{I}_{\N}(t)^{-2} I(U_i\geqslant t>L_i)^2 \big\{I(X_i\geqslant t>L_i)-g(\Beta_{\D}^{t\trans}\Phibar_{\D i}^{t})\big\}^2,\\\nonumber
     \text{s.t.} & \quad \frac{1}{n}\sum_{i=1}^{n} I(U_i\geqslant t>L_i)\big\{ I(X_i\geqslant t>L_i)- g(\Beta_{\D}^{t\trans}\Phibar_{\D i}^{t})\big\}=0,
\end{align}
where $\widebar{I}_{\N}(t) = \frac{1}{N}\sum_{i=n+1}^{n+N}I(U_i\geqslant t > L_i)$. 
We can show that $\widehat{\Beta}_{\ID}^{t}$ is a consistent estimator of 
\begin{align}
\label{dintrexp}
    \widebar{\Beta}_{\ID}^{t}&= \arg\min\limits_{\Beta_{\D}^{t}} \,I_{\D}(t)^{-2} E\bigl[I(U\geqslant t>L)\big\{I(X\geqslant t >L)-g(\Beta_{\D}^{t\trans}\Phibar_{\D}^{t})\big\}\bigr]^{2},\\ \nonumber
    & \text{s.t.} \quad E\bigl[I(U \geqslant t > L) \big\{ I(X \geqslant t > L)- g(\Beta_{\D}^{t\trans}\Phibar_{\D}^{t})\big\}\bigr]=0.
\end{align}
We then estimate the survival rate $S(t)$ by marginalizing imputed risk $g(\widehat{\Beta}_{\ID}^{t\trans}\Phibar_{\D i}^{t})$ of unlabeled data $\mathcal{D}_u$ as follows,
\begin{equation}
    \label{dinst}
    \Shat_\ssd=\frac{\sum_{i=n+1}^{n+N}I(U_i\geqslant t>L_i)g(\widehat{\Beta}_{\ID}^{t\trans}\Phibar_{\D i}^{t})}{\sum_{i=n+1}^{n+N}I(U_i\geqslant t>L_i)}.
\end{equation}
The moment condition in \eqref{dintreq} is used to calibrate the potential bias from a misspecified imputation model and ensure the consistency of the survival estimator $\Shat_\ssd$. The following theorem gives the asymptotic properties of $\Shat_\ssd$. The proofs and regularity conditions needed are presented in Appendix B.
\begin{theorem}
\label{themID}
Under Conditions 3-6 given in the Appendix, $\Shat_\ssd \stackrel{p}{\longrightarrow} S(t)$, and
\begin{align*}
&\sqrt{n}\left(\Shat_\ssd-S(t)\right)\\
=\,\,\,& \frac{1}{\sqrt{n}} \sum_{i=1}^{n} I_{\D}(t)^{-1} I(U_i\geqslant t>L_i)\big\{ I(X_i\geqslant t>L_i)- g(\widebar{\Beta}_{\ID}^{t\trans}\Phibar_{\D i}^{t})\big\} + o_p(1),
\end{align*}
which converges to $N\big(0, \Sigma_\ssd \big)$ in distribution, where
\begin{equation}
\label{dintrasyvar}
\Sigma_\ssd =I_{\D}(t)^{-2}E\bigl[I(U\geqslant t>L)\big\{I(X\geqslant t>L)-g(\widebar{\Beta}_{\ID}^{t\trans}\Phibar_{\D}^{t})\big\}\bigr]^{2},
\end{equation}
which can be consistently estimated by
\begin{equation*}
    \Sighat_\ssd=\frac{1}{n}\sum_{i=1}^{n} \widebar{I}_{\N}(t)^{-2} I(U_i\geqslant t>L_i)^2\big\{I(X_i\geqslant t>L_i)- g(\widehat{\Beta}_{\ID}^{t\trans}\Phibar_{\D i}^{t})\big\}^2.
\end{equation*}
\end{theorem}
By the definition of $\widebar{\Beta}_{\D}^t$ and $\widebar{\Beta}_{\ID}^{t}$, comparing the asymptotic variances in \eqref{dsslasyvar} and \eqref{dintrasyvar}, one can easily see that the intrinsic estimator $\Shat_\ssd$ is more efficient than $\Shat_{\D}(t)$, especially when the imputation model is misspecified. 

For comparison, we present the supervised estimator that only uses the labeled data $\mathcal{D}_{l}$.
Note that $T$ and $(L, U)$ are independent, we have ${\rm Pr}(X\geqslant t>L)={\rm Pr}(U\geqslant t >L, T\geqslant t)={\rm Pr}(U\geqslant t>L)\>{\rm Pr}(T\geqslant t)$.
Hence, similar to the imputation method in \cite{CaiCheng2004}, we construct a supervised estimator by using labeled data $\mathcal{D}_{l}$ as follows,
\begin{equation}
\label{dslst}
    \Shat_\sd=\frac{\sum_{i=1}^{n}I(X_i\geqslant t>L_i)}{\sum_{i=1}^{n}I(U_i\geqslant t>L_i)}.
\end{equation}
As shown in the Appendix B, $\sqrt{n}\>\big(\Shat_\sd - S(t)\big)$ converges to a zero-mean Gaussian distribution with asymptotic variance
\begin{equation}
\label{dslasyvar}
    \Sigma_\sd=I_{\D}(t)^{-2}E\left[I(U \geqslant t > L) \big\{I(X\geqslant t>L) - S(t)\big\}\right]^{2},
\end{equation}
which can be consistently estimated by
\begin{equation*}
    \Sighat_\sd = \frac{1}{n}\sum_{i=1}^{n}\widebar{I}_{n}(t)^{-2} I(U_i\geqslant t>L_i)^2\big\{I(X_i\geqslant t>L_i)- \Shat_\sd\big\}^2,
\end{equation*}
where $\bar{I}_n(t) = \frac{1}{n}\sum_{i=1}^{n}I(U_i\geqslant t>L_i)$.

Comparing the asymptotic variances in \eqref{dintrasyvar} and \eqref{dslasyvar}, one can see that the semi-supervised estimator $\Shat_\ssd$ gains efficiency over the supervised estimator $\Shat_\sd$ by using the additional information in unlabeled data $\mathcal{D}_{u}$ when model \eqref{dmodel} is correctly specified.

\subsubsection{Estimation based on the left (right) censoring label}
\label{sect22}

Here we develop a semi-supervised estimator of the survival function $S(t)$ by using the left censoring label $\big\{L_i, \, I(T_i < L_i)\big\}$. Estimation using the right censoring label $\big\{U_i, \,I(T_i > U_i)\big\}$ is similar and thus omitted. In this section, with a little abuse of notation, we still use $\mathcal{D}_{l}$ to denote the labeled data and redefine it as $\mathcal{D}_{l}=\bigl\{\bigl(L_i, I(T_i< L_i),\mathbf{X}_{i}^{\ast \trans},\bm{\delta}_{i}^{\ast\trans},\bfZ_i^{\trans},\mathbf{\widebar{Z}}_{Li}^{\trans}\bigr)^{\trans},i=1,\ldots,n\bigr\}$. Since $T$ is independent of $L$, we have
\begin{align*}
S(t) = {\rm Pr}(T\geqslant t) = E\big[I(T \geqslant L)\mid L=t\big] = E\Big[E\big[I(T\geqslant L) \mid \widebar{\textbf{W}}_{\L}^{t},L=t\big]\mid L=t\Big],
\end{align*}
where $\widebar{\textbf{W}}_{\L}^{t} = \big(1, \textbf{W}_{\L}^{t\trans}\big)^{\trans}$ and $\textbf{W}_{\L}^{t}=\big(\mathbf{X}^{\ast\trans},\bm{\delta}^{\ast\trans},\bfZ^{\trans},\mathbf{Z}_{L}^{t\trans}\big)^{\trans}$.

First, we use labeled data $\mathcal{D}_{l}$ to fit the following working time-specific logistic model
\begin{equation}
\label{lmodel}
E\big[I(T\geqslant L)\mid  \widebar{\textbf{W}}_{\L}^{t},L=t\big]=g(\Beta_{\L}^{t\trans}\Phibar_{\L}^{t}),
\end{equation}
where $\Beta_{\L}^{t}$ is an unknown time-specific regression parameter, $\Phibar_{\L}^{t} = \big(1, \bm\Phi(\textbf{W}_{\L}^{t})^{\trans}\big)^{\trans}$ for some known basis function $\bm\Phi(\cdot)$, and $g(x)=\exp(x)/(1$ $+\exp(x))$.
Similar to \cite{Ahuja2021semi}, we obtain an estimator of $\Beta_{\L}^{t}$, denoted by $\widehat{\Beta}_{\L}^{t}$, by solving the following estimating equation,
\begin{equation}
\label{lest}
    \mathbf{U}_n(\Beta_{\L}^{t})=\frac{1}{n}\sum_{i=1}^{n}K_{h_l}(L_i-t)\Phibar_{\L i}^{t}\big\{I(T_i\geqslant L_i)-g(\Beta_{\L}^{t\trans}\Phibar_{\L i}^{t})\big\}=0,
\end{equation}
where $K_{h_l}(\cdot)=K(\cdot/h_{l})/h_{l}$, $K(\cdot)$ is a kernel density function, $h_{l}$ is the bandwidth, and $h_{l} = O(n^{-\kappa_1})$ with $\kappa_1 \in (1/5, 1/2)$. Under mild conditions, we can show that $\widehat{\Beta}_{\L}^{t}$ is a consistent estimator of the unique solution $\widebar{\Beta}_{\L}^t$ to $\widebar{\mathbf{U}}(\Beta_{\L}^{t})=f_l(t)E\bigl[\Phibar_{\L}^{t} \bigl\{I(T\geqslant t)-g(\Beta_{\L}^{t\trans}\Phibar_{\L}^{t})\bigr\}\bigr]=0$, where $f_l(t)$ is the density function of left censoring variable $L$.

Next, with $\widehat{\Beta}_{\L}^{t}$ in place, we estimate the survival function $S(t)$ by marginalizing imputed risks $g(\widehat{\Beta}_{\L}^{t\trans}\Phibar_{\L i}^{t})$ of unlabeled data $\mathcal{D}_u$ as below, 
$$
\Shat_{\L}(t)=\frac{\sum_{i=n+1}^{n+N}K_{h_{\L}}(L_i-t)g(\widehat{\Beta}_{\L}^{t\trans}\Phibar_{\L i}^{t})}{\sum_{i=n+1}^{n+N}K_{h_{\L}}(L_i-t)},
$$
where $K_{h_{\L}}(\cdot)=K(\cdot/h_{\L})/h_{\L}$, $h_{\L}$ is the bandwidth, and $h_{\L} = O(N^{-\kappa_2})$ with $\kappa_2 \in (1/5,1/2)$.
In Appendix B, we show that  $\sqrt{nh_l}\big(\Shat_{\L}(t)-S(t)\big)$ converges to a normal distribution with mean zero and asymptotic variance
\begin{equation}
\label{lsslasyvar}
    \Sigma_{\L}(t)=\nu_2 f_l(t)^{-1}E\bigl[I(T\geqslant t) - g(\widebar{\Beta}_{\L}^{t\trans}\Phibar_{\L}^{t})\bigr]^2,
\end{equation}
where $\nu_2 = \int K^{2}(u)\diff u$. 

Lastly, similar to Section~\ref{sect21}, we improve $\Shat_{\L}(t)$ using the idea of intrinsic efficiency by finding an estimator of $\Beta_{\L}^{t}$ that asymptotically minimizes the variance \eqref{lsslasyvar}. Specifically, we obtain the estimator of $\Beta_{\L}^{t}$ as
\begin{align}
\label{Lintreq}
    \widehat{\Beta}_{\IL}^{t}=& \arg\min\limits_{\Beta_{\L}^{t}}\, \frac{h_{l}}{n}\sum_{i=1}^{n} \widebar{K}_\NL(t)^{-2} K_{h_l}^2(L_i-t)\big\{I(T_i\geqslant L_i)-g(\Beta_{\L}^{t\trans}\Phibar_{\L i}^{t})\big\}^2,\\\nonumber
    \text{s.t.} & \quad \frac{1}{n}\sum_{i=1}^{n} K_{h_l}(L_i-t)\big\{I(T_i\geqslant L_i)-g(\Beta_{\L}^{t\trans}\Phibar_{\L i}^{t})\big\}=0,
\end{align}
where $\widebar{K}_\NL(t) = \frac{1}{N}\sum_{i=n+1}^{n+N}K_{h_{\L}}(L_i-t)$ converges to $f_{l}(t)$ as $h_{\L}\to 0$, $N \to \infty$, and $Nh_{\L}\to \infty$. It can be shown that $\widehat{\Beta}_{\IL}^{t}$ converges to
\begin{align}
    \label{Lintrexp}
    &\widebar{\Beta}_{\IL}^{t}= \arg\min\limits_{\Beta_{\L}^{t}}\, \nu_2 f_{l}(t)^{-1} E\bigl[I(T \geqslant t)- g(\Beta_{\L}^{t\trans}\Phibar_{\L}^{t})\bigr]^{2},\\\nonumber
    & \text{s.t.} \quad  f_{l}(t) E\bigl[I(T \geqslant t)- g(\Beta_{\L}^{t\trans}\Phibar_{\L}^{t})\bigr]=0.
\end{align}
Then by marginalizing the imputed risk $g(\widehat{\Beta}_{\IL}^{t\trans}\Phibar_{\L i}^{t})$ from unlabeled data $\mathcal{D}_u$, we obtain the following estimator of $S(t)$, 
\begin{equation}
\label{intrSL}
\Shat_\ssl=\frac{\sum_{i=n+1}^{n+N}K_{h_{\L}}(L_i-t)g(\widehat{\Beta}_{\IL}^{t\trans}\Phibar_{\L i}^{t})}{\sum_{i=n+1}^{n+N}K_{h_{\L}}(L_i-t)}.
\end{equation}
The asymptotic properties of $\Shat_\ssl$ are described in the following theorem. The regularity conditions and proofs are presented in Appendix B.
\begin{theorem}
\label{themIL}
Under Conditions 1-6 given in the Appendix, we have $\Shat_\ssl \stackrel{p}{\longrightarrow} S(t)$, and
\begin{align*}
\sqrt{nh_{l}}\left(\Shat_\ssl-S(t)\right) = \sqrt{\frac{h_{l}}{n}} \sum_{i=1}^{n} f_{l}(t)^{-1} K_{h_l}(L_i-t) \big\{ I(T_i\geqslant L_i)- g(\widebar{\Beta}_{\IL}^{t\trans}\Phibar_{\L i}^{t})\big\} + o_p(1),
\end{align*}
which converges to $N\big(0, \Sigma_\ssl\big)$ in distribution, where
\begin{equation}
\label{lintrasyvar}
    \Sigma_\ssl= \nu_2 f_l(t)^{-1} E\bigl[I(T \geqslant t )-g(\widebar{\Beta}_{\IL}^{t\trans}\Phibar_{\L}^{t})\bigr]^2,
\end{equation}
which can be consistently estimated by
\begin{equation*}
    \Sighat_\ssl=\frac{h_{l}}{n}\sum_{i=1}^{n} \widebar{K}_\NL(t)^{-2} K_{h_l}^2(L_i-t)\big\{I(T_i\geqslant L_i) - g(\widehat{\Beta}_{\IL}^{t\trans}\Phibar_{\L i}^{t})\big\}^2.
\end{equation*}
\end{theorem}

To make a comparison, we present the supervised estimator of $S(t)$ below that only uses labeled data $\mathcal{D}_{l}$. Since $T$ and $L$ are independent, we have $S(t) = {\rm Pr}(T \geqslant t) = {\rm Pr}( T\geqslant L \mid L=t)$. Then following the IPCW method proposed by \cite{Van1998locally}, we obtain the supervised estimator
\begin{equation}
\label{lslst}
    \Shat_\sl=\frac{\sum_{i=1}^{n}K_{h_l}(L_i - t) I(T_i \geqslant L_i)}{\sum_{i=1}^{n}K_{h_l}(L_i - t)}.
\end{equation}
In Appendix B, we show that under mild regularity conditions, $\sqrt{nh_{l}}\>\big(\Shat_\sl - S(t)\big)$ converges to a zero-mean Gaussian distribution with asymptotic variance
\begin{equation}
\label{lslasyvar}
    \Sigma_\sl = \nu_2 f_l(t)^{-1} E\left[I(T_i \geqslant t) - S(t)\right]^{2},
\end{equation}
which can be consistently estimated by
\begin{equation*}
    \Sighat_\sl=\frac{h_{l}}{n}\sum_{i=1}^{n} \widebar{K}_{nl}(t)^{-2} K_{h_l}^{2}(L_i -t)\bigl\{I(T_i \geqslant L_i) - \Shat_\sl \bigr\}^2,
\end{equation*}
where $\widebar{K}_{nl}(t) = \frac{1}{n}\sum_{i=1}^{n}K_{h_l}(L_i - t)$, which goes to $f_l(t)$ as $h_l\to 0$, $n\to \infty$, and $nh_l\to \infty$. One can see from the asymptotic variances in \eqref{lintrasyvar} and \eqref{lslasyvar} that the supervised estimator $\Shat_\sl$ is not as efficient as the semi-supervised estimator $\Shat_\ssl$ that additionally uses the surrogate and covariate information in the unlabeled data $\mathcal{D}_{u}$ when model \eqref{lmodel} is correctly specified.

Estimation based on the right-censoring label $\big\{U_i, \,I(T_i > U_i)\big\}$ can be constructed similarly. Let $\Shat_\su$ and $\Shat_\ssu$ denote the corresponding supervised and semi-supervised estimators, respectively. Also, we denote their corresponding asymptotic variances by $\Sigma_\su$ and $\Sigma_\ssu$ with estimators $\Sighat_\su$ and $\Sighat_\ssu$. So far, in Step 1 of our proposed estimation procedure, we have obtained three semi-supervised estimators based on three types of labels. In Step 2 following, we develop our final SEEDS estimator as an optimal linear combination of the three estimators from Step 1.

\subsection{Step 2: Combination of three estimators}

In Step 2, we develop a combined semi-supervised estimator, called SEEDS, as an optimal linear combination of the three semi-supervised estimators $\mathbf{\Shat}_\SS =\big(\Shat_\ssd, \Shat_\ssl, \Shat_\ssu\big)^{\trans}$ obtained from Step 1. This estimator is optimal in the sense that it has the smallest variance among all linear combinations of $\mathbf{\Shat}_\SS$ where the linear coefficient sums up to 1, and thus is consistent and more efficient than any of the three estimators. More details are given below. For comparison, using the same idea, we construct a combined supervised estimator, called CSL, as an optimal linear combination of the three supervised estimators $\mathbf{\Shat}_\S =\big(\Shat_\sd, \Shat_\sl, \Shat_\su\big)^{\trans}$.

\subsubsection{Proposed SEEDS estimator}

As mentioned above, we construct our SEEDS estimator as an optimal linear combination of the three semi-supervised estimators $\mathbf{\Shat}_\SS =\big(\Shat_\ssd, \Shat_\ssl, \Shat_\ssu\big)^{\trans}$ from Step 1. In particular, we define the SEEDS estimator of $S(t)$ as 
\begin{equation}
    \label{cosslst}
    \Shat_\ssc=\Mhat_\SS^{\trans}\mathbf{\Shat}_\SS,
\end{equation}
where $\Mhat_\SS=\bm{\Vhat}_\SS^{-1}\mathbf{1}/(\mathbf{1}^{\trans}\bm{\Vhat}_\SS^{-1}\mathbf{1})$ and $\bm{\Vhat}_\SS$ is a consistent estimator of the covariance matrix of $\mathbf{\Shat}_\SS$. Specifically, the diagonal elements of $\bm{\Vhat}_\SS$ are given by $\Sighat_\ssd / n$, $\Sighat_\ssl / (n h_l)$, and $\Sighat_\ssu / (n h_u)$, where $\Sighat_\ssd$ is defined in Section~\ref{sect21}, $\Sighat_\ssl$ is defined in Section~\ref{sect22}, and $\Sighat_\ssu$, $h_u$, and $h_{\U}$ are defined similarly as $\Sighat_\ssl$, $h_l$, and $h_{\L}$ based on the right censoring label instead. Note that $\bm{\Vhat}_\SS$ is a symmetric matrix. The off-diagonal elements of $\bm{\Vhat}_\SS$, denoted by $\Vhat_{\mbox{\tiny SSDL}}(t)$, $\Vhat_{\mbox{\tiny SSDU}}(t)$, and $\Vhat_{\mbox{\tiny SSLU}}(t)$, are given below. The covariance estimator of $\Shat_\ssd$ and $\Shat_\ssl$ is given by
\begin{align*}
    \Vhat_{\mbox{\tiny SSDL}}(t) = &\frac{1}{n^2}\sum_{i=1}^{n}I(U_i\geqslant t>L_i)\big\{I(X_i\geqslant t>L_i) - g(\widehat{\Beta}_{\ID}^{t\trans}\Phibar_{\D i}^{t} )\big\}K_{h_l}(L_i - t)\big\{I(T_i \geqslant L_i) -\\
    &g(\widehat{\Beta}_{\IL}^{t\trans}\Phibar_{\L i}^{t})\big\} \big/ \big\{ \widebar{I}_{\N}(t) \widebar{K}_{\NL}(t)\big\},
\end{align*}
the covariance estimator of $\Shat_\ssd$ and $\Shat_\ssu$ is
\begin{align*}
    \Vhat_{\mbox{\tiny SSDU}}(t) = &\frac{1}{n^2}\sum_{i=1}^{n}I(U_i\geqslant t>L_i)\big\{I(X_i\geqslant t>L_i) - g(\widehat{\Beta}_{\ID}^{t\trans}\Phibar_{\D i}^{t} )\big\}K_{h_u}(U_i - t) \big\{I(T_i \geqslant U_i) -\\
    &g(\widehat{\Beta}_{\IU}^{t\trans}\Phibar_{\U i}^{t})\big\} \big/ \big\{\widebar{I}_{\N}(t) \widebar{K}_{\NU}(t)\big\},
\end{align*}
and the covariance estimator of $\Shat_\ssl$ and $\Shat_\ssu$ is
\begin{align*}
    \Vhat_{\mbox{\tiny SSLU}}(t) = &\frac{1}{n^2}\sum_{i=1}^{n}K_{h_l}(L_i - t) \big\{I(T_i \geqslant L_i) - g(\widehat{\Beta}_{\IL}^{t\trans}\Phibar_{\L i}^{t})\big\}K_{h_u}(U_i - t)\big\{I(T_i \geqslant U_i) -g(\widehat{\Beta}_{\IU}^{t\trans}\Phibar_{\U i}^{t})\big\}\\
    &\big/ \big\{\widebar{K}_{\NL}(t) \widebar{K}_{\NU}(t)\big\},
\end{align*}
where $\widebar{K}_{\NL}(t)$ is defined in Section~\ref{sect22} and $\widebar{K}_{\NU}(t)$ is similarly defined based on the right censoring label instead.

From Theorem~\ref{themID} and Theorem~\ref{themIL}, it is easy to see that the proposed SEEDS estimator $\Shat_\ssc$ is consistent and asymptotically normal. 
Also, the asymptotic variance of $\Shat_\ssc$ can be consistently estimated by
\begin{align}
\label{comssvar}
    \Vhat_\ssc = &\frac{1}{n^2} \sum_{i = 1}^{n} \Big[ \mhat_{\mbox{\tiny ss1}}(t) \widebar{I}_{\N}(t)^{-1} I(U_i\geqslant t>L_i)\big\{I(X_i\geqslant t>L_i)- g(\widehat{\Beta}_{\ID}^{t\trans}\Phibar_{\D i}^{t} )\big\}\\\nonumber
    & + \mhat_{\mbox{\tiny ss2}}(t)\widebar{K}_{\NL}(t)^{-1}K_{h_l}(L_i - t) \big\{I(T_i \geqslant L_i) - g(\widehat{\Beta}_{\IL}^{t\trans}\Phibar_{\L i}^{t})\big\} \\\nonumber
    &+ \mhat_{\mbox{\tiny ss3}}(t)\widebar{K}_{\NU}(t)^{-1} K_{h_u}(U_i - t) \big\{I(T_i \geqslant U_i)-g(\widehat{\Beta}_{\IU}^{t\trans}\Phibar_{\U i}^{t})\big\}\Big]^2,
\end{align}
where $\mhat_{\mbox{\tiny ssj}}(t), j=1,2,3$ are the elements of $\Mhat_\SS$. The SEEDS estimator is more efficient than each of the three semi-supervised estimators $\mathbf{\Shat}_\SS =\big(\Shat_\ssd, \Shat_\ssl, \Shat_\ssu\big)^{\trans}$ from Step 1.

\subsubsection{Combined supervised (CSL) estimator}

For comparison, we construct the supervised counterpart of our SEEDS estimator by combining the three supervised estimators $\mathbf{\Shat}_\S=\bigl(\Shat_\sd$, $\Shat_\sl, \Shat_\su\bigr)^{\trans}$ given in Section~\ref{sect21} and Section~\ref{sect22}. The combined supervised (CSL) estimator of $S(t)$ is defined as
\begin{equation}
    \label{coslst}
    \Shat_\sc=\Mhat_\S^{\trans}\mathbf{\Shat}_\S,
\end{equation}
where $\Mhat_\S=\bm{\Vhat}_\S^{-1}\mathbf{1}/(\mathbf{1}^{\trans}\bm{\Vhat}_\S^{-1}\mathbf{1})$ and $\bm{\Vhat}_\S$ is a consistent estimator of the covariance matrix with the diagonal elements $\Sighat_\sd / n$, $\Sighat_\sl / (n h_l)$, and $\Sighat_\su / (n h_u)$ given by Section~\ref{sect21} and Section~\ref{sect22}, and with the off-diagonal elements $\Vhat_{\mbox{\tiny SDL}}(t)$, $\Vhat_{\mbox{\tiny SDU}}(t)$, and $\Vhat_{\mbox{\tiny SLU}}(t)$ given below. The covariance estimator of $\Shat_\sd$ and $\Shat_\sl$ is
\begin{align*}
    \Vhat_{\mbox{\tiny SDL}}(t) = & \frac{1}{n^2}\sum_{i=1}^{n}I(U_i\geqslant t>L_i)\big\{I(X_i\geqslant t>L_i) - \Shat_\sd\big\}K_{h_l}(L_i - t)\big\{I(T_i \geqslant L_i) - \Shat_\sl\big\} \\
    &\big/ \big\{\widebar{I}_n(t) \widebar{K}_{nl}(t)\big\},
\end{align*}
the covariance estimator of $\Shat_\sd$ and $\Shat_\su$ is
\begin{align*}
    \Vhat_{\mbox{\tiny SDU}}(t) = &\frac{1}{n^2}\sum_{i=1}^{n}I(U_i\geqslant t > L_i)\big\{I(X_i\geqslant t > L_i) - \Shat_\sd\big\}K_{h_u}(U_i - t)\big\{I(T_i \geqslant U_i) - \Shat_\su\big\} \\
    &\big/ \big\{\widebar{I}_n(t) \widebar{K}_{nu}(t)\big\},
\end{align*}
and the covariance estimator of $\Shat_\sl$ and $\Shat_\su$ is
\begin{align*}
    \Vhat_{\mbox{\tiny SLU}}(t) = &\frac{1}{n^2}\sum_{i=1}^{n}K_{h_l}(L_i - t) \big\{I(T_i \geqslant L_i) - \Shat_\sl\big\}K_{h_u}(U_i - t)\big\{I(T_i \geqslant U_i) - \Shat_\su\big\} \\
    &\big / \big\{\widebar{K}_{nl}(t) \widebar{K}_{nu}(t)\big\},
\end{align*}
where $\widebar{K}_{nl}(t)$ is defined in Section~\ref{sect22} and $\widebar{K}_{nu}(t)$ is similarly defined based on the right censoring label instead. 

Given the results of supervised estimators in Section~\ref{sect21} and Section~\ref{sect22}, it is easy to show that the CSL estimator $\Shat_\sc$ is consistent and asymptotically normal, and the asymptotic variance of  $\Shat_\sc$ can be consistently estimated by
\begin{align*}
    \Vhat_\sc = &\frac{1}{n^2} \sum_{i = 1}^{n} \Big[\mhat_{\mbox{\tiny s1}}(t) \widebar{I}_{n}(t)^{-1} I(U_i\geqslant t>L_i)\big\{I(X_i\geqslant t>L_i)- \Shat_\sd\big\} + \mhat_{\mbox{\tiny s2}}(t)\widebar{K}_{nl}(t)^{-1}K_{h_l}(\\
    &L_i- t) \big\{I(T_i \geqslant L_i) - \Shat_\sl\big\}+ \mhat_{\mbox{\tiny s3}}(t)\widebar{K}_{nu}(t)^{-1} K_{h_u}(U_i - t) \big\{I(T_i \geqslant U_i)- \Shat_\su\big\}\Big]^2,
\end{align*}
where $\mhat_{\mbox{\tiny sj}}(t), j=1,2,3$ are the elements of $\Mhat_\S$. As with the comparison of supervised and semi-supervised estimators in Section~\ref{sect21} and Section~\ref{sect22}, the CSL estimator is not as efficient as the SEEDS estimator that additionally uses the unlabeled data when the imputation models are correctly specified.

\subsection{Cross-fitting method}
\label{secsslcv}

When calculating the SEEDS estimator from \eqref{cosslst}, the plug-in covariance matrix estimator $\bm{\Vhat}_\SS$ reuses the labeled data and may thus underestimate the true variance-covariance in small samples due to over-fitting. In this section, we develop a $K$-fold cross-validation (CV) method to correct for such bias \citep{Efron1986bias}. In particular, we randomly split the labeled data of size $n$ into $K$ folds $\mathcal{I}_{1}, \ldots,$ $\mathcal{I}_{\textit{K}}$ of roughly equal size. We first use $K-1$ folds $\mathcal{I}_{\k} = \{1, \ldots, n\} / \mathcal{I}_{\textit{k}}$ to estimate $\Beta_{\dob}^{t}$ and denote the estimator by $\widehat{\Beta}_{\dob,\k}^{t}$, where $\Beta_{\dob}^{t}$ generically represents $\Beta_{\ID}^{t}$, $\Beta_{\IL}^{t}$, and $\Beta_{\IU}^{t}$ corresponding to the three types of labeled data discussed above. We then use the remaining part $\mathcal{I}_{k}$ to estimate the covariance matrix by plugging-in $\widehat{\Beta}_{\dob,\k}^{t}$. More details are given below. 

First, using $K-1$ folds of the labeled data $\mathcal{I}_{\k} = \{1, \ldots, n\} / \mathcal{I}_{\textit{k}}$, we obtain the estimators $\widehat{\Beta}_{\ID,\k}^{t}$ and $\widehat{\Beta}_{\IL,\k}^{t}$ from \eqref{dintreq} and \eqref{Lintreq}, respectively. $\widehat{\Beta}_{\IU,\k}^{t}$ can be calculated similarly as $\widehat{\Beta}_{\IL,\k}^{t}$ using the right censoring label instead. Then we use the remaining fold $\mathcal{I}_{\textit{k}}$ to estimate the covariance matrix of $\mathbf{\Shat}_\SS =\big(\Shat_\ssd, \Shat_\ssl, \Shat_\ssu\big)^{\trans}$ by plugging-in $\widehat{\Beta}_{\dob,\k}^{t}$. Repeating this procedure for $k=1,\ldots,K$, we obtain the CV-calibrated covariance matrix estimator $\bm{\Vhat}^{\cv}_\SS$ as follows. The diagonal elements of $\bm{\Vhat}^{\cv}_\SS$, that is, CV-calibrated variance estimators of $\Shat_\ssd, \Shat_\ssl$ and $\Shat_\ssu$ are given by
\begin{align*}
    \label{ssvarcv} 
 \Vhat^{\cv}_\ssd =\frac{1}{K} \sum_{k = 1}^{K} \frac{1}{|\mathcal{I}_{k}|} \sum_{i \in I_{k}} \widebar{I}_{\N}(t)^{-2} I(U_i\geqslant t>L_i)^2\big\{I(X_i\geqslant t>L_i)- g(\widehat{\Beta}_{\ID,\k}^{t\trans}\Phibar_{\D i}^{t} )\big\}^2,\\
    \Vhat^{\cv}_\ssl = \frac{1}{K} \sum_{k = 1}^{K} \frac{1}{|\mathcal{I}_{k}|} \sum_{i \in I_{k}}\widebar{K}_\NL(t)^{-2}  K_{h_l}^2(L_i-t)\big\{I(T_i\geqslant L_i) - g(\widehat{\Beta}_{\IL,\k}^{t\trans}\Phibar_{\L i}^{t})\big\}^2,\\
    \Vhat^{\cv}_\ssu = \frac{1}{K} \sum_{k = 1}^{K} \frac{1}{|\mathcal{I}_{k}|} \sum_{i \in I_{k}}\widebar{K}_\NU(t)^{-2} K_{h_u}^2(U_i-t)\big\{I(T_i\geqslant U_i) - g(\widehat{\Beta}_{\IU,\k}^{t\trans}\Phibar_{\U i}^{t})\big\}^2.
\end{align*}
Note that $\bm{\Vhat}^{\cv}_\SS$ is a $3\times3$ symmetric matrix. The off-diagonal elements of $\bm{\Vhat}^{\cv}_\SS$, that is, CV-calibrated covariance estimators of $\Shat_\ssd, \Shat_\ssl$ and $\Shat_\ssu$ are given by
\begin{align*}
    \Vhat_{\mbox{\tiny SSDL}}^{\cv}(t) = &\frac{1}{K} \sum_{k = 1}^{K} \frac{1}{|\mathcal{I}_{k}|} \sum_{i \in I_{k}} I(U_i\geqslant t>L_i)\big\{I(X_i\geqslant t>L_i) - g(\widehat{\Beta}_{\ID,\k}^{t\trans}\Phibar_{\D i}^{t})\big\}K_{h_l}(L_i - t)\\ \nonumber
    &\big\{I(T_i \geqslant L_i) - g(\widehat{\Beta}_{\IL,\k}^{t\trans}\Phibar_{\L i}^{t})\big\} \big/ \big\{ \widebar{I}_{\N}(t) \widebar{K}_{\NL}(t)\big\},\\
    \Vhat_{\mbox{\tiny SSDU}}^{\cv}(t) = &\frac{1}{K} \sum_{k = 1}^{K} \frac{1}{|\mathcal{I}_{k}|} \sum_{i \in I_{k}} I(U_i\geqslant t>L_i)\big\{I(X_i\geqslant t>L_i) - g(\widehat{\Beta}_{\ID,\k}^{t\trans}\Phibar_{\D i}^{t})\big\}K_{h_u}(U_i - t) \\ 
    &\big\{I(T_i \geqslant U_i) - g(\widehat{\Beta}_{\IU,\k}^{t\trans}\Phibar_{\U i}^{t})\big\} \big/ \big\{ \widebar{I}_{\N}(t) \widebar{K}_{\NU}(t)\big\},\\
    \Vhat_{\mbox{\tiny SSLU}}^{\cv}(t) = &\frac{1}{K} \sum_{k = 1}^{K} \frac{1}{|\mathcal{I}_{k}|} \sum_{i \in I_{k}} K_{h_l}(L_i - t) \big\{I(T_i \geqslant L_i) - g(\widehat{\Beta}_{\IL,\k}^{t\trans}\Phibar_{\L i}^{t})\big\}K_{h_u}(U_i - t) \big\{I(T_i \geqslant U_i)\\
    &-g(\widehat{\Beta}_{\IU,\k}^{t\trans}\Phibar_{\U i}^{t})\big\} \big/ \big\{\widebar{K}_{\NL}(t) \widebar{K}_{\NU}(t)\big\}.
\end{align*}
With the CV-calibrated covariance estimator $\bm{\Vhat}^{\cv}_\SS$, we obtain our SEEDS estimator as 
\begin{equation}
    \label{comcvssl}
    \Shat^{\cv}_\ssc=\Mhat^{\cv}_\SS^{\trans} \> \mathbf{\Shat}_\SS,
\end{equation}
where $\Mhat^{\cv}_\SS=\bm{\Vhat}^{\cv}_\SS^{-1}\mathbf{1}/(\mathbf{1}^{\trans}\bm{\Vhat}^{\cv}_\SS^{-1}\mathbf{1})$. The variance of $\Shat^{\cv}_\ssc$ can be estimated by
\begin{align*}
    \Vhat^{\cv}_\ssc = &\frac{1}{K} \sum_{k = 1}^{K} \frac{1}{|\mathcal{I}_{k}|} \sum_{i \in I_{k}} \Big[ \mhat_{\mbox{\tiny ss1}}^{\cv}(t) \widebar{I}_{\N}(t)^{-1} I(U_i\geqslant t>L_i)\big\{I(X_i\geqslant t>L_i)- g(\widehat{\Beta}_{\ID,\k}^{t\trans}\Phibar_{\D i}^{t})\big\} \\
    &+ \mhat_{\mbox{\tiny ss2}}^{\cv}(t) \widebar{K}_{\NL}(t)^{-1} K_{h_l}(L_i - t) \big\{I(T_i \geqslant L_i) - g(\widehat{\Beta}_{\IL,\k}^{t\trans}\Phibar_{\L i}^{t})\big\}\\
    &+ \mhat_{\mbox{\tiny ss3}}^{\cv}(t) \widebar{K}_{\NU}(t)^{-1} K_{h_u}(U_i - t) \big\{I(T_i \geqslant U_i)- g(\widehat{\Beta}_{\IU,\k}^{t\trans}\Phibar_{\U i}^{t})\big\}\Big]^2.
\end{align*}
where $\mhat^{\cv}_{\mbox{\tiny ssj}}(t), j=1,2,3$ are the elements of $\Mhat^{\cv}_\SS$. In practice, $\bm{\Vhat}^{\cv}_\SS$ may be unstable due to high correlation of $\Shat_\ssd$, $\Shat_\ssl$, and $\Shat_\ssu$. One could use the regularized estimator $\big(\bm{\Vhat}^{\cv}_\SS + \delta_n \mathbf{I}\big)^{-1}$ for some $\delta_n = O(n^{-\frac{1}{2}})$ to stabilize the estimation and obtain $\Shat_\ssc$ accordingly.

\section{Simulation Study}
\label{simul}

In this section, we conducted extensive simulation studies to evaluate the finite-sample performance of our proposed SEEDS estimator and also compare it with the CSL estimator that only uses labeled data. We considered various data-generating mechanisms. In particular, we generated the surrogate $T^{\ast}$ from a uniform distribution, the baseline covariate $Z$ from a normal distribution, the survival time $T$ from the Cox model or logistic model conditional on $T^{\ast}$ and $Z$, the left censoring variable $L$ from a Weibull distribution, the right censoring variable $U$ from a uniform distribution plus $L$, and the time-dependent covariate $\widebar{Z}_{L}^{t}$ from a Poisson process whose mean rate depends on $T$. Various settings with or without $Z$ or $\widebar{Z}_{L}^{t}$ are considered. Here we only present simulation results from the following two settings, where both baseline covariate $Z$ and time-dependent covariate $\widebar{Z}_{L}^{t}$ are included and the correlation between $T$ and $T^{\ast}$ is around $0.67$. More results from other settings can be found in Appendix A.
\begin{itemize}
  \item[1.] $T^{\ast}\sim {\rm Un}(0, 1/2)$, \\
 $Z \sim {\rm Norm}(5, 1)$, \\
 ${\rm Pr}(T \geqslant t \mid T^{\ast}, Z) = \exp\{-t^{3} \exp(-7.6 T^\ast - 0.15 Z) / 0.6\}$, \\
 $\widebar{Z}_{L}^{t}\sim {\rm Poisson\> Process}(\lambda(T))$ with $\lambda(T) = 2 T$, \\
 $L\sim {\rm Weibull}(1.38, 1.3)$, $U \sim L+{\rm Un}(0, 3.3)$,\\
  yielding 27\% left-censoring rate and 29\% right-censoring rate.
 \item[2.] $T^{\ast}\sim {\rm Un}(-1, 1)$,\\
  $Z \sim {\rm Norm}(5, 1)$, \\
  ${\rm Pr}(T \geqslant t \mid T^{\ast}, Z) = 1/[1+\exp\{(t - 2 - 0.95 T^{\ast} - 0.1 Z)/0.33\}]$, \\
  $\widebar{Z}_{L}^{t}\sim {\rm Poisson\> Process}(\lambda(T))$ with $\lambda(T) = T$, \\
  $L\sim {\rm Weibull}(2.1, 1.95)$, $U \sim L+{\rm Un}(0, 3)$, \\
  yielding 25\% left-censoring rate and 32\% right-censoring rate.
\end{itemize}
For all the settings considered, we used the Gaussian kernel with bandwidth from a rule of thumb \citep{Wand1990kernel} and satisfying the required undersmoothing and we employed the 10-fold cross-validation procedure to correct for the over-fitting bias. We evaluated the proposed SEEDS estimator and the CSL estimator of $S(t)$ at 50 equally spaced grid points from 10$\%$ to 90$\%$ quantiles of the observed times. For each setting, we calculated the empirical biases, empirical standard errors (ESE), average of the estimated standard errors (ASE), empirical coverage probabilities (CovP) of the 95$\%$ confidence intervals for $S(t)$, and relative efficiencies (RE) with respect to mean squared errors compared to the CSL estimator based on 500 simulated datasets. The sizes for labeled and unlabeled data are taken as 250 and 5000, respectively.

\begin{figure}[htbp]
\centerline{\includegraphics[height=4.5in,width=5.5in]{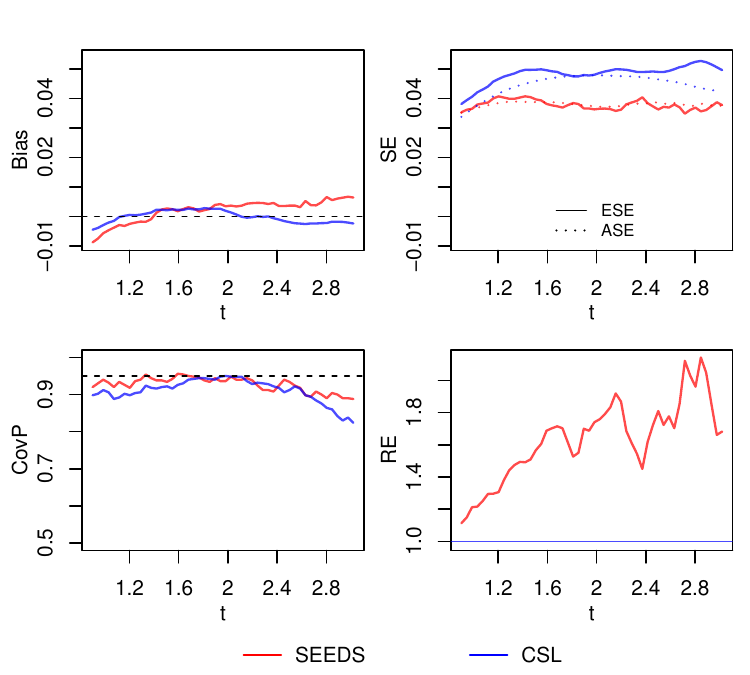}}
\caption{Empirical biases, empirical standard errors (ESE), averages of the estimated standard errors (ASE), empirical coverage probabilities (CovP) of the 95$\%$ confidence intervals for the SEEDS and the CSL estimators, and relative efficiency (RE) of the proposed SEEDS estimator compared to the CSL estimator with respect to mean squared errors, respectively, under setting 1.}
\label{figure1}
\end{figure}

\begin{figure}[htbp]
\centerline{\includegraphics[height=4.5in,width=5.5in]{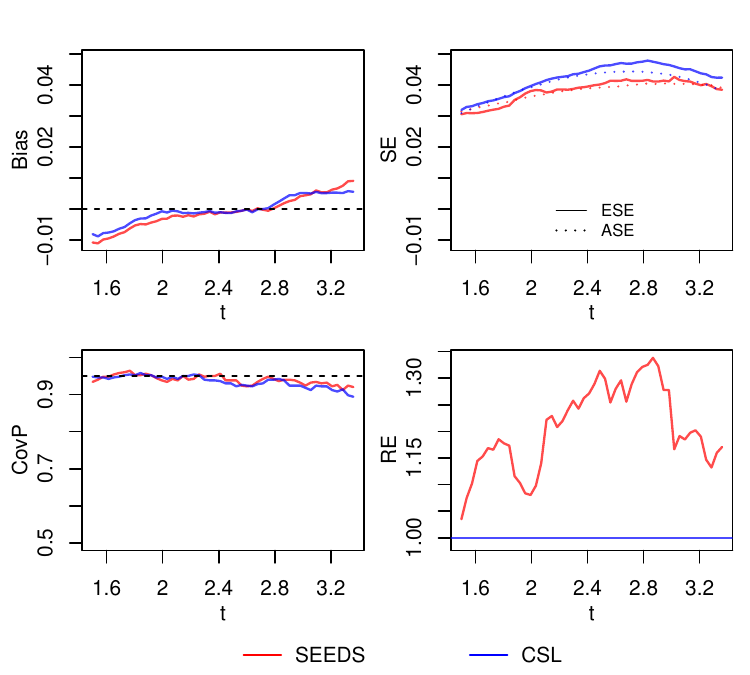}}
\caption{Empirical biases, empirical standard errors (ESE), averages of the estimated standard errors (ASE), empirical coverage probabilities (CovP) of the 95$\%$ confidence intervals for the SEEDS and the CSL estimators, and relative efficiency (RE) of the proposed SEEDS estimator compared to the CSL estimator with respect to mean squared errors, respectively, under setting 2.}
\label{figure2}
\end{figure}

Results for the two above settings are presented in Figure~\ref{figure1} and Figure~\ref{figure2}, respectively. We can see that for both SEEDS and CSL estimators, the biases are negligible, standard error estimates reflect true variability, and the coverage probabilities are close to the nominal level except a bit off at the tails due to kernel estimation. The SEEDS method performs well in both settings where $T$ is generated from the Cox model and logistic model, respectively, implying that it is robust to the misspecification of the imputation model. Also, as expected, by leveraging the surrogate and covariate information from unlabeled data, the SEEDS estimator is more efficient than the CSL estimator with relative efficiency as high as 2 at some time points. Similar conclusions are observed for the other settings considered in Appendix A.

\section{Application to T2D Patients from MGB}
\label{applic}

We applied the proposed SEEDS method to EHR data on Type 2 Diabetes (T2D) from the Massachusetts General Brigham (MGB) health system. The dataset consists of $115,236$ patients who have at least one International Classification of Diseases (ICD) code \citep{Liao2010electronic, Cipparone2015inaccuracy}, including $64,808$ females and $50,428$ males. The event time $T$ of interest is age at the T2D onset. Some patients have already developed T2D at the first visit, leading to left-censored observations, while some others have not yet developed T2D at the last visit, yielding right-censored observations. Thus, the event time $T$ is subject to double censoring. Specifically, we defined the left censoring time $L$ as age at the first visit and defined the right censoring time $U$ as age at the last visit in the health system thus far. The true outcome of T2D onset time was ascertained by manual chart review only for a small labeled sample of $1,613$ patients, including $915$ females and $698$ males. Among the $1,613$ labeled patients, the event time $T$ is exactly observed for $168\,(10.4\%)$ patients, right-censored for $1,401\,(86.9\%)$ patients, and left-censored for $44\,(2.7\%)$ patients. For females, these percentages are $8.6\%$, $90.3\%$, and $1.1\%$, respectively. For males, they are $12.8\%$, $82.4\%$, and $4.9\%$, respectively. In addition, we defined age at the first encounter of PheCode for T2D (PheCode:250.2) as a surrogate outcome $T^\ast$ which is also doubly-censored by $L$ and $U$. If $T^\ast < L$, it is a left-censored observation; if $T^\ast>U$, it is a right-censored observation. Among all the $115,236$ patients in the dataset, the surrogate outcome $T^\ast$ is exactly observed for $16,987\,(14.7\%)$ patients, right-censored for $95,597\,(83\%)$ patients, and left-censored for $2,652\,(2.3\%)$ patients. For females, these percentages are $12.7\%$, $85.8\%$, and $1.5\%$, respectively. For males, these percentages are $17.3\%$, $79.3\%$, and $3.4\%$, respectively.

\begin{figure}[htbp]
	\centerline{\includegraphics[height=3.5in,width=6.5in]{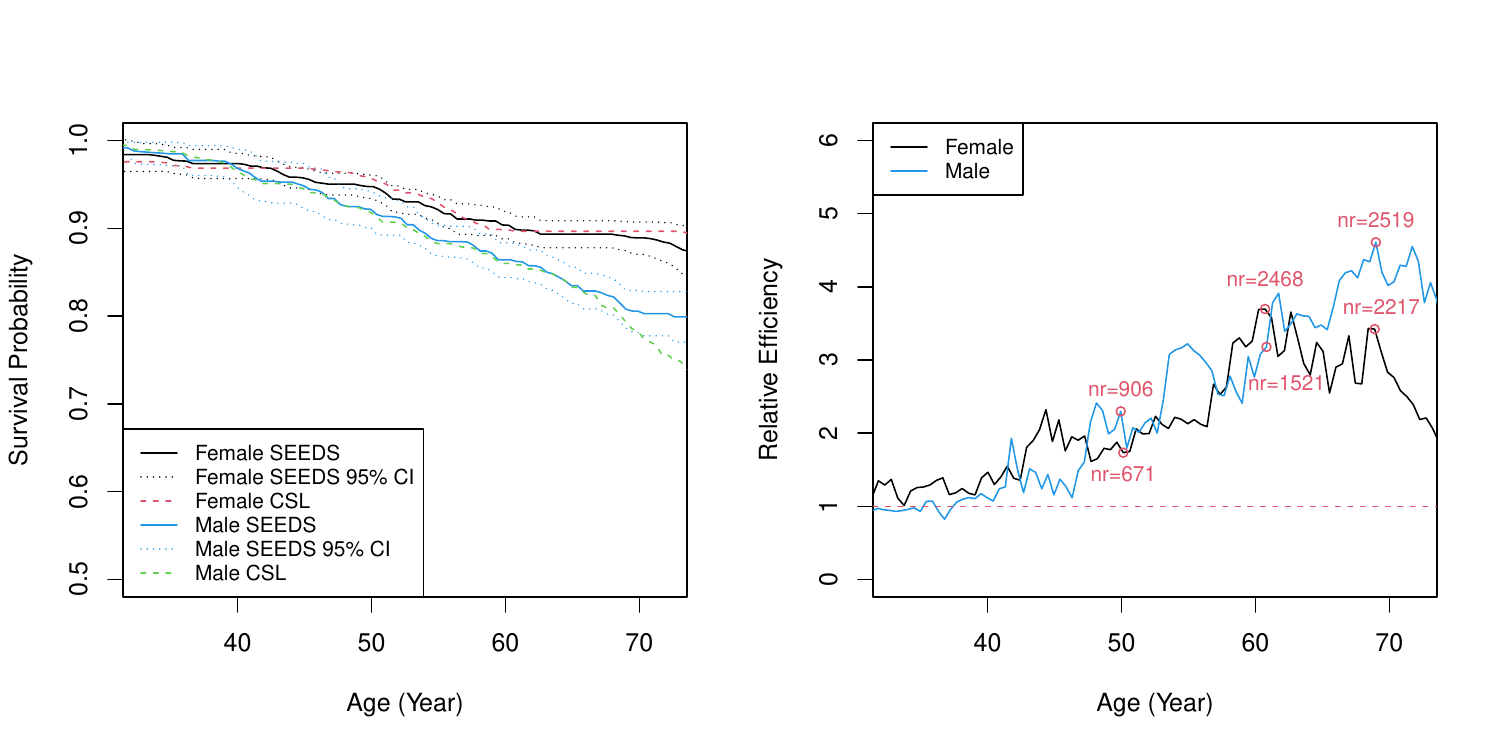}}
	\caption{On the left panel, the survival estimates using the SEEDS and CSL methods, respectively, together with the 95$\%$ confidence interval for the SEEDS method are presented. On the right panel, the relative efficiency of SEEDS to CSL based on their estimated variances is presented, and the sample size of additional labeled data required for CSL to achieve the same efficiency as SEEDS, denoted by $nr$, is provided.}
	\label{figure3}
\end{figure}

We employed the proposed SEEDS method to estimate the survival functions of female and male groups, respectively, and also obtained the CSL estimates for comparison. The estimation results are presented in Figure~\ref{figure3}. One can see from the results that the survival (T2D-free) probabilities are generally higher for females than for males and the difference gets larger over time. Also, the proposed SEEDS estimator gains a lot of efficiency over the CSL estimator and the efficiency gain seems to get larger at later times. In particular, the relative efficiency of SEEDS to CSL is as high as $3.7$ for females at year $61$ and as high as $4.6$ for males at year $69$. In Figure~\ref{figure3}, we also report the sample size of additional labeled data required for the CSL estimator to achieve the same efficiency as the SEEDS estimator. For example, at age 69 in males, an additional set of $2519$ labels would be needed for the CSL method to achieve the same precision as the SEEDS method in estimating the survival rate. It highlights the effectiveness of our SEEDS method in leveraging the information from unlabeled data and thereby reducing the cost of acquiring labels by labor-intensive manual chart review.

\section{Discussion}
\label{dissc}

We have developed a robust estimation procedure of survival rate based on doubly-censored EHR data that includes a small set of labeled data with gold-standard survival outcomes captured by manual chart review and a large set of unlabeled data containing proxies or surrogate outcomes. By leveraging the information in unlabeled data, the proposed semi-supervised SEEDS method achieves great efficiency gain over the supervised CSL method that only uses labeled data. It thus considerably reduces the burden of labor-intensive manual chart review and improves the feasibility of using EHR data to study disease risk. We have shown that the proposed SEEDS estimator is consistent regardless of whether the imputation model holds, while the magnitude of efficiency gain of the SEEDS estimator over the CSL estimator depends on the accuracy of imputed risks, which relates to the association of proxies or surrogates with 
the true disease as well as the fit of imputation model. In addition, we have developed a cross-fitting method to correct for the potential over-fitting bias in small samples due to repeated use of labeled data.

There are some extensions or directions for future research. The proposed estimation procedure assumes that the left censoring time $L$ and the right censoring time $U$ are independent of covariates. This assumption may not be applicable in practice. The imputation-based strategy developed here can be extended to handle the situation where $L$ and $U$ depend on covariates. However, this is not trivial and definitely warrants future research. On the other hand, risk prediction is an important goal in clinical practice. Developing risk prediction models, as well as methods for evaluating predictive accuracy based on doubly-censored EHR data, warrants future research.

\section*{Acknowledgements}

We gratefully acknowledge support of the Shanghai Jiao Tong University Research Project ``Outstanding Doctoral Graduate Development Grant". 
\vspace*{-8pt}

\section*{Data Availability Statement}

The findings in this paper are supported by data available from Mass General Brigham Health System. However, access to these data is subject to restrictions as they were used under a licensing agreement for this paper. Interested individuals can obtain the data from the authors with the permission of Mass General Brigham.
\vspace*{-8pt}


\section*{Appendix}

 Appendix A includes some additional simulation results. Appendix B provides proof of our theoretical results. 

\section{Additional Simulation Results}

In this section, we present additional simulation results under various settings. In particular, besides the scenario in the main paper, we consider three scenarios regarding whether time-independent or time-dependent covariates are available besides the surrogate $T^{\ast}$. For each scenario, we consider different data-generating mechanisms and the correlation between $T^{\ast}$ and $T$ is around 0.67. In total, 6 settings are considered and the details for each setting are given below. The results are presented in Figures 1$\sim$6, respectively. We can see that for both the SEEDS and CSL estimators, the biases are negligible, standard error estimates reflect true variability, and the coverage probabilities are close to the nominal level. The SEEDS estimator is generally more efficient than the CSL estimator, especially when time-dependent covariates are available and included in the imputation models.

\begin{itemize}
\item[S1.] Scenarios with surrogate and no other covariates
   \begin{itemize}
       \item[1.1] $T^{\ast}\sim {\rm Un}(0, 2/3)$, ${\rm Pr}(T \geqslant t \mid T^{\ast}) = \exp\{-t^{12/5}\exp(-5.85\, T^\ast) / 0.3\}$, $L\sim {\rm Weibull}$ $(1.05, 0.72)$, $U \sim L+{\rm Un}(0, 1.6)$, $X^{\ast}=\max(L,\min(T^{\ast},U))$, yielding 25\% left-censoring rate, 42\% right-censoring rate.

       \item[1.2] $T^{\ast}\sim {\rm Un}(-1, 1)$, ${\rm Pr}(T \geqslant t \mid T^{\ast}) = 1/[1+\exp\{(t - 2 - 0.95\, T^{\ast})/0.35\}]$, $L\sim {\rm Weibull}(2.5, 2.45)$, $U \sim L+{\rm Un}(0, 2.25)$, $X^{\ast}=\max(L,\min(T^{\ast},U))$, yielding 25\% left-censoring rate, 42\% right-censoring rate.
       
   \end{itemize}
   
\item[S2.] Scenarios with surrogate and time-independent covariates
   \begin{itemize}
       \item[2.1] $T^{\ast}\sim {\rm Un}(0, 2/3)$, $Z \sim {\rm Norm}(5, 1)$, ${\rm Pr}(T \geqslant t \mid T^{\ast}, Z) = \exp\{-t^{12/5}\exp(-6.5\, T^\ast - 0.4\, Z) / 0.15\}$, $L\sim {\rm Weibull}(0.98, 1.4)$, $U \sim L+{\rm Un}(0, 3)$, $X^{\ast}=\max(L,\min(T^{\ast},U))$, yielding 27\% left-censoring rate, 42\% right-censoring rate.
       
       \item[2.2] $T^{\ast}\sim {\rm Un}(-1, 1)$, $Z \sim {\rm Norm}(5, 1)$, ${\rm Pr}(T \geqslant t \mid T^{\ast}, Z) = 1/[1+\exp\{(t - 2 - 0.95\, T^{\ast} - 0.1\, Z)/0.33\}]$, $L\sim {\rm Weibull}(2.1, 1.95)$, $U \sim L+{\rm Un}(0, 3)$, $X^{\ast}=\max(L,\min(T^{\ast},U))$, yielding 25\% left-censoring rate, 32\% right-censoring rate.
       
   \end{itemize}
   
\item[S3.] Scenarios with surrogate and time-dependent covariates 
\begin{itemize}
   \item[3.1] $T^{\ast}\sim {\rm Un}(0, 2/3)$, ${\rm Pr}(T \geqslant t \mid T^{\ast}) = \exp\{- t^{12/5} \exp(-6.1\, T^\ast) / 0.3\}$, $\widebar{Z}_{\L}^{t}\sim {\rm Poisson}$ ${\rm Process} (\lambda(T))$, and $\lambda(T) = 2\, T$, $L\sim {\rm Weibull}(1.04, 0.75)$, $U \sim L+{\rm Un}(0, 1.4)$, $X^{\ast}=\max(L,\min(T^{\ast},U))$, yielding 25.6\% left-censoring rate, 46\% right-censoring rate.
   
   \item[3.2] $T^{\ast}\sim {\rm Un}(-1, 1)$, ${\rm Pr}(T \geqslant t \mid T^{\ast}) = 1/[1+\exp\{(t - 3 - T^{\ast})/0.35\}]$, $\widebar{Z}_{\L}^{t}\sim{\rm Poisson}\,{\rm Process}(\lambda(T))$, and $\lambda(T) = T$, $L\sim {\rm Weibull}(2.5, 2.45)$, $U \sim L+{\rm Un}(0, 2.25)$, $X^{\ast}=\max(L,\min(T^{\ast},U))$, yielding 25.4\% left-censoring rate, 42.4\% right-censoring rate.
  \end{itemize}
 \end{itemize}

\begin{figure}[htbp]
\centerline{\includegraphics[height=3in,width=4in]{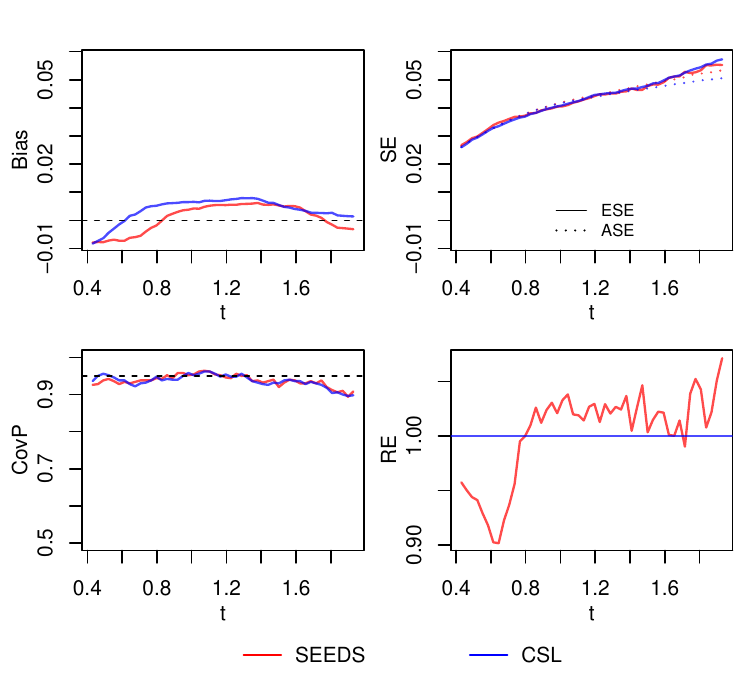}}
\caption{Empirical biases, empirical standard errors (ESE), averages of the estimated standard errors (ASE), empirical coverage probabilities (CovP) of the 95$\%$ confidence intervals for the SEEDS and the CSL estimators, relative efficiency (RE) of the proposed SEEDS estimator corresponding to the CSL estimator with respect to mean squared errors, respectively, under setting 1.1.}
\label{figure1-1}
\end{figure}

\begin{figure}[htbp]
\centerline{\includegraphics[height=3in,width=4in]{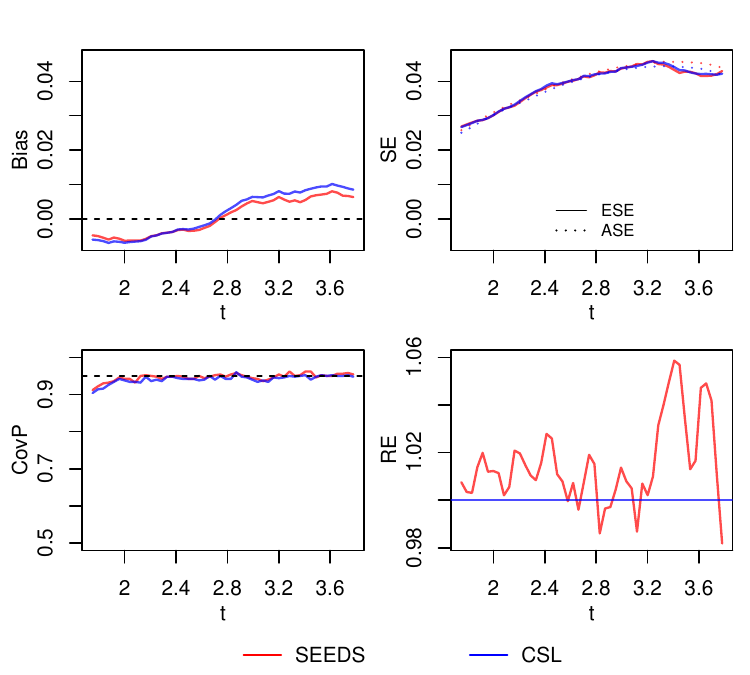}}
\caption{Empirical biases, empirical standard errors (ESE), averages of the estimated standard errors (ASE), empirical coverage probabilities (CovP) of the 95$\%$ confidence intervals for the SEEDS and the CSL estimators, relative efficiency (RE) of the proposed SEEDS estimator corresponding to the CSL estimator with respect to mean squared errors, respectively, under setting 1.2.}
\label{figure1-2}
\end{figure}

\begin{figure}[htbp]
\centerline{\includegraphics[height=3in,width=4in]{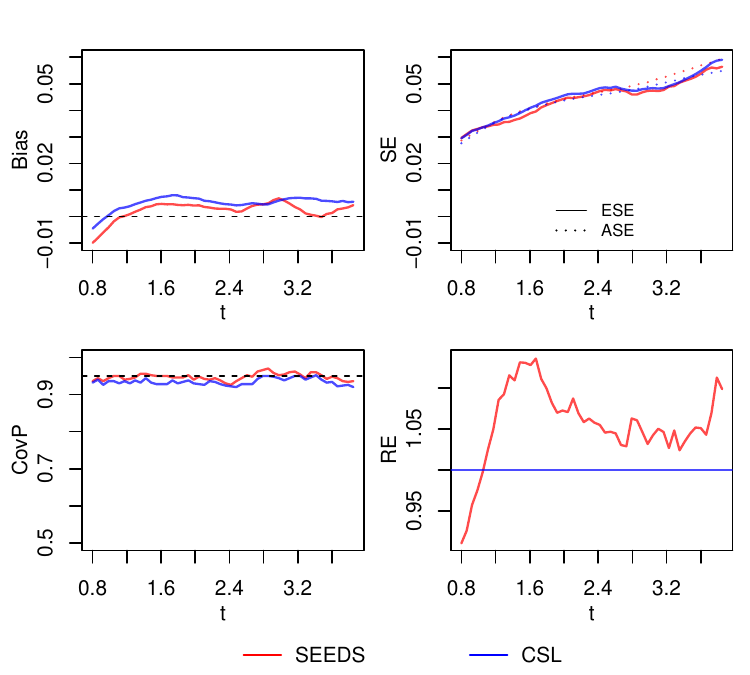}}
\caption{Empirical biases, empirical standard errors (ESE), averages of the estimated standard errors (ASE), empirical coverage probabilities (CovP) of the 95$\%$ confidence intervals for the SEEDS and the CSL estimators, relative efficiency (RE) of the proposed SEEDS estimator corresponding to the CSL estimator with respect to mean squared errors, respectively, under setting 2.1.}
\label{figure2-1}
\end{figure}

\begin{figure}[htbp]
\centerline{\includegraphics[height=3in,width=4in]{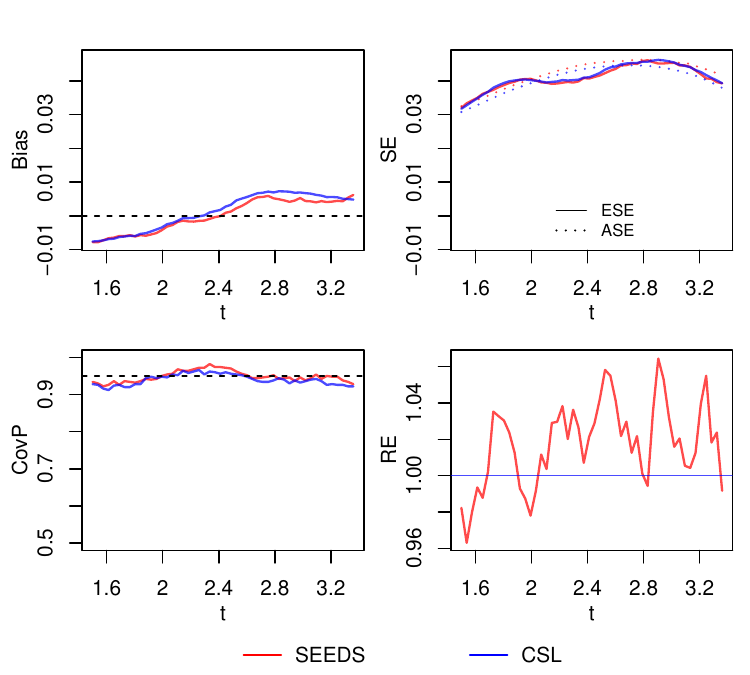}}
\caption{Empirical biases, empirical standard errors (ESE), averages of the estimated standard errors (ASE), empirical coverage probabilities (CovP) of the 95$\%$ confidence intervals for the SEEDS and the CSL estimators, relative efficiency (RE) of the proposed SEEDS estimator corresponding to the CSL estimator with respect to mean squared errors, respectively, under setting 2.2.}
\label{figure2-2}
\end{figure}

\begin{figure}[htbp]
\centerline{\includegraphics[height=3in,width=4in]{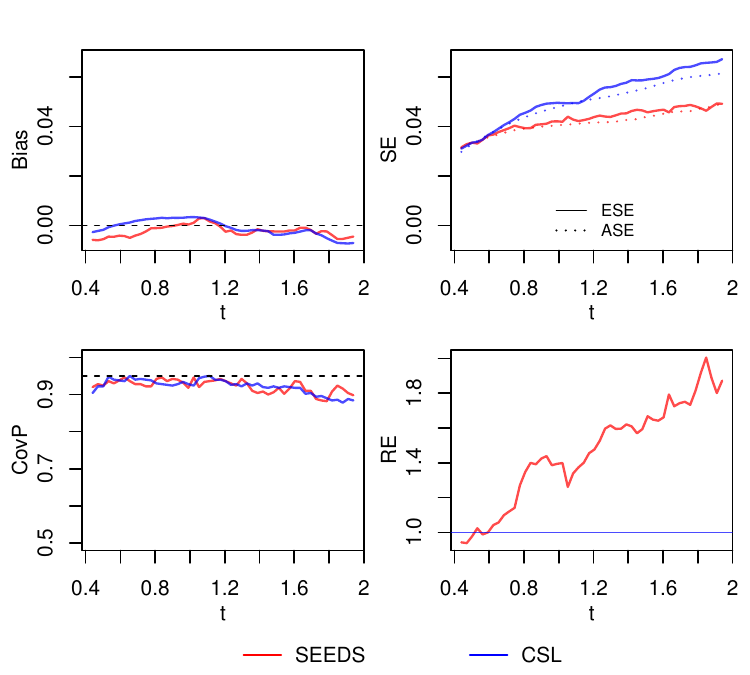}}
\caption{Empirical biases, empirical standard errors (ESE), averages of the estimated standard errors (ASE), empirical coverage probabilities (CovP) of the 95$\%$ confidence intervals for the SEEDS and the CSL estimators, relative efficiency (RE) of the proposed SEEDS estimator corresponding to the CSL estimator with respect to mean squared errors, respectively, under setting 3.1.}
\label{figure3-1}
\end{figure}

\begin{figure}[htbp]
\centerline{\includegraphics[height=3in,width=4in]{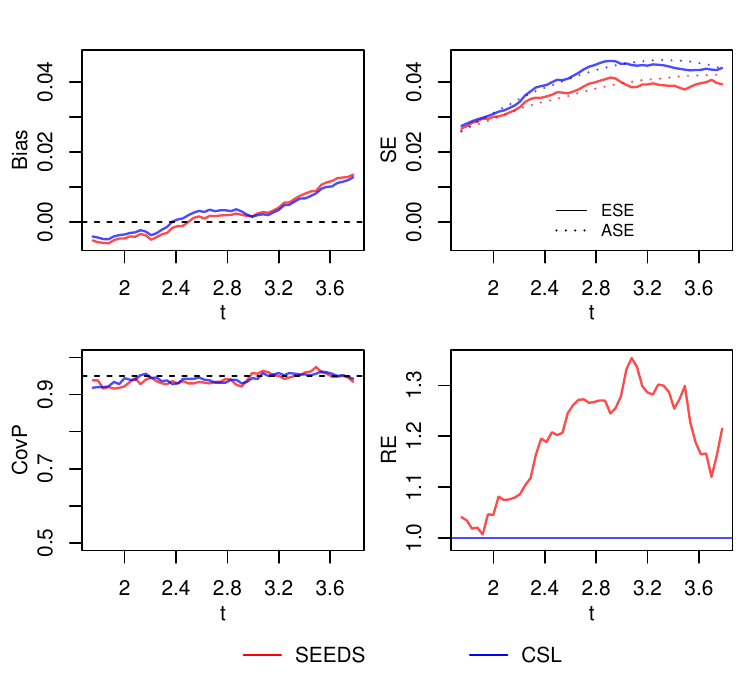}}
\caption{Empirical biases, empirical standard errors (ESE), averages of the estimated standard errors (ASE), empirical coverage probabilities (CovP) of the 95$\%$ confidence intervals for the SEEDS and the CSL estimators, relative efficiency (RE) of the proposed SEEDS estimator corresponding to the CSL estimator with respect to mean squared errors, respectively, under setting 3.2.}
\label{figure3-2}
\end{figure}

\newpage
\section{Proofs of the Asymptotic Properties}

In this section, we provide proof of our theoretical results. Define $\bm\Sigma_1 \succ \bm\Sigma_2$ if $\bm\Sigma_1 - \bm\Sigma_2$ is positive definite for any two symmetric matrices $\bm \Sigma_1$ and $\bm\Sigma_2$, and define $\textbf{v}^{\otimes 2} = \textbf{v} \textbf{v}^{\trans}$ for any vector $\textbf{v}$. Let $\mathcal{T} = [\tau_a, \tau_b]$ for some constants $\tau_a$ and $\tau_b$ satisfying $0<\tau_a < \tau_b<\infty$,  $Y_{\mbox{\tiny X}} = I(X \geqslant t >L)$, and $Y_{\mbox{\tiny T}} = I(T\geqslant t)$. Define
\begin{align*}
    \textbf{A}_1 = & I_{\D}(t)^{-2}E\Big[ I\left(U \geqslant t > L\right)^2 \big[\ddot{g}(\widebar{\Beta}_{\ID}^{t\trans} \Phibar_{\D}^{t}) \big\{I\left(X \geqslant t > L\right)-g(\widebar{\Beta}_{\ID}^{t\trans} \Phibar_{\D}^{t})\big\} -\dot{g}^{2}(\widebar{\Beta}_{\ID}^{t\trans}\Phibar_{\D}^{t})\big]\Phibar_{\D}^{t\otimes 2} \Big];\\
    \textbf{A}_2 = & \nu_2 f_{l}(t)^{-1} E\Big[\big[\ddot{g}(\widebar{\Beta}_{\IL}^{t\trans} \Phibar_{\L}^{t})\big\{I\left(T \geqslant t\right)-g(\widebar{\Beta}_{\IL}^{t \trans} \Phibar_{\L}^{t})\big\}-\dot{g}^{2}(\widebar{\Beta}_{\IL}^{t\trans} \Phibar_{\L}^{t})\big]\Phibar_{\L}^{t\otimes 2}\Big];\\
    \textbf{B}_1 = &E\left[I\left(U \geqslant t > L \right)\dot{g}(\widebar{\Beta}_{\ID}^{t\trans} \Phibar_{\D}^{t})\Phibar_{\D}^{t} \right]; \\
    \textbf{B}_2 =&f_l(t)E\left[\dot{g}(\widebar{\Beta}_{\IL}^{t\trans} \Phibar_{\L}^{t})\Phibar_{\L}^{t} \right].
\end{align*}
Also, define $\mathcal{B}_{\epsilon \dob}=\{\bm{\beta_{\cdot}}^{t}: \|\bm{\beta_{\cdot}}^{t} - \widebar{\Beta}_{\dob}^{t}\|_2 < \epsilon \dob\}$. Note that we use the subscript $\dob$ to generically represent the notations for three types of labels. The regularity conditions needed for the theoretical results are given below.
\begin{itemize}
    \item[C1] The kernel function $K(\cdot) \geqslant 0$ is a symmetric density function with compact support $\mathcal{K} \subseteq \mathbb{R}$, and is of bounded variation on $\mathcal{K}$;
    \item[C2] The bandwidth $h_l = O(n^{-\kappa_1})$ for $\kappa_1 \in (1/5, 1/2)$ and $h_{\L} = O(N^{-\kappa_2})$ for $\kappa_2 \in (1/5, 1/2)$; 
    \item[C3] $\log{N}/\log{n}\to \kappa_0>8/5$, as $n\to \infty$;
    \item[C4] The density functions of $L$ and $U$ are $f_l(t)$ and $f_u(t)$, respectively. They are continuous and satisfy $\inf_{t \in \mathcal{T}} f_l(t) > 0$, $\inf_{t \in \mathcal{T}} f_u(t) > 0$, and ${\rm Pr}(L < U) =1$;
    \item[C5] The basis function ${\bm\Phi}(\textbf{W}_{\dob}^{t})$ contains $\textbf{W}_{\dob}^{t}$, has compact support, and is of fixed dimension. The density function of $\textbf{W}_{\dob}^{t}$, denoted by $p(\textbf{W}_{\dob}^{t})$, is continuously differentiable in the continuous components of $\textbf{W}_{\dob}^{t}$;
    \item[C6] There is no vector $\bm{\beta_{\cdot}}^{t}$ such that ${\rm Pr}(Y_{\dob 1} > Y_{\dob 2} \mid \Beta^{t \trans}_{\dob 1} \Phibar_{\dob 1}^{t} > \Beta^{t \trans}_{\dob 2} \Phibar_{\dob 2}^{t}) = 1$ for all $t \in \mathcal{T}$. Also, $E\big[\Phibar_{\dob}^{t\otimes 2} \dot{g}(\widebar{\Beta}^{t \trans}_{\dob} \Phibar_{\dob}^{t})\big] \succ 0$, $\textbf{A}_1 \succ 0$, $\textbf{A}_2 \succ 0$, $\textbf{B}_1^{\trans} \textbf{A}_1^{-1} \textbf{B}_1 \succ 0$, and $\textbf{B}_2^{\trans} \textbf{A}_2^{-1} \textbf{B}_2 \succ 0$.
\end{itemize}
Conditions C1, C2, C4, and C5 are fairly standard in the literature. Condition C3 enlists some fundamental characteristics of the semi-supervised setting. Similar to \citet{tian2007model}, Condition C6 ensures the existence and uniqueness of the limiting parameters $\Beta_{\dob}^{t}$.

First, we give the following lemma, which is important for proving our theoretical results.
\begin{lemma}
\label{lem1}
Let 
$$C_n(t, \Beta_{\dob}^{t}) = \frac{1}{n}\sum_{i=1}^{n}\Phibar_{\dob i}^{t}K_h(L_i -t) g(\Beta^{t \trans}_{\dob}\Phibar_{\dob i}^{t})$$
and assume Conditions C1-C5 hold, then 
$$\sup_{t \in \mathcal{T}, \Beta_{\dob} \in \mathcal{B}_{\epsilon \dob}}\left\|C_n(t, \Beta^{t}_{\dob})- C(t, \Beta^{t}_{\dob}) \right\|_2 = o_p(1),$$
where $C(t, \Beta^{t}_{\dob}) = E\big[\Phibar_{\dob}^{t}g(\Beta^{t \trans}_{\dob}\Phibar_{\dob}^{t})\big]f_l(t)$.
\end{lemma}
\textbf{Proof} Since $K(\cdot)$ is of bounded variation, from the result of Example 2.10 in \citet{pakes1989simulation}, we obtain $\{K((\cdot - t)/h), t \in \mathcal{T}\}$ is Euclidean. Under Condition C5, we know $\Phibar_{\dob}^{t}$ has compact support and is of fixed dimension, and $g(\cdot)$ is expit function, then $g(\Beta^{t \trans}_{\dob}\Phibar_{\dob i}^{t})$ is of bounded variation. Thus, we can write $g(\Beta^{t \trans}_{\dob}\Phibar_{\dob i}^{t}) = g_1(\Beta^{t \trans}_{\dob}\Phibar_{\dob i}^{t})-g_2(\Beta^{t \trans}_{\dob}\Phibar_{\dob i}^{t})$, where both $g_1(\Beta^{t \trans}_{\dob}\Phibar_{\dob i}^{t})$ and $g_2(\Beta^{t \trans}_{\dob}\Phibar_{\dob i}^{t})$ are nonnegative and nondecreasing. Using the arguments employed in Lemma A2 of \citet{bilias1997towards}, we find both $g_1(\Beta^{t \trans}_{\dob}\Phibar_{\dob i}^{t})$ and $g_2(\Beta^{t \trans}_{\dob}\Phibar_{\dob i}^{t})$ are Euclidean. Therefore, by Lemma 2.14 of \citet{pakes1989simulation}, $\bigl\{\Phibar_{\dob i}^{t}K_h(L_i -t) g(\Beta^{t \trans}_{\dob}\Phibar_{\dob i}^{t}), t\in \mathcal{T}, \Beta^{t}_{\dob} \in \mathcal{B}_{\epsilon \dob}\bigr\}$ is Euclidean. Hence it must be manageable according to Theorem 4.8 of \citet{pollard1990empirical}. Choose envelopes as $B_1/\sqrt{n}$, for some constant $B_1$. Then we have $\sup_{t \in \mathcal{T}, \Beta_{\dob} \in \mathcal{B}_{\epsilon \dob}}\big\|C_n(t, \Beta^{t}_{\dob})- C(t, \Beta^{t}_{\dob}) \big\|_2 = o_p(1)$ by Theorem 8.2 (ULLN) of \citet{pollard1990empirical}. The proof is completed.

\subsection{Asymptotic properties of the estimators based on the exact observed event time label}

In this section, we establish the asymptotic properties of the supervised estimator $\Shat_\sd$, semi-supervised estimator $\Shat_{\D}$, intrinsic semi-supervised estimator $\Shat_\ssd$, respectively, based on the exact observed event time label.

\subsubsection{Asymptotic properties of the supervised estimator $\Shat_\sd$}
\label{Dslpf}
Here we derive the asymptotic properties of the supervised estimator $\Shat_\sd$. First, we prove consistency. Note that 
\begin{align*}
    \Shat_\sd& = \sum_{i=1}^{n}I(X_i\geqslant t>L_i) \big/ \sum_{i=1}^{n}I(U_i\geqslant t>L_i)\\
    & = E\left[I(X_i \geqslant t > L_i)\right] \big/ E\left[I(U_i \geqslant t > L_i)\right]+ o_p(1)\\
    & = E\left[I(U_i \geqslant t > L_i) I(T_i \geqslant t)\right] \big/ E\left[I(U_i \geqslant t > L_i)\right]+ o_p(1)\\
    & = S(t) + o_p(1).
\end{align*}
Under the above mild regularity conditions, it follows from Theorem 8.2 (ULLN) of \citet{pollard1990empirical} that $\Shat_\sd \stackrel{p}{\longrightarrow} S(t)$ uniformly in $t\in \mathcal{T}$.

Next, we consider the asymptotic normality,
\begin{align}
\label{dsupasy}
    \sqrt{n}\>\left(\Shat_\sd - S(t)\right) = &\frac{1}{\sqrt{n}}\sum_{i=1}^{n} \bar{I}_n(t)^{-1} I(U_i\geqslant t>L_i)\big\{I(X_i\geqslant t>L_i) - S(t)\big\}\\ \nonumber
    = & \frac{1}{\sqrt{n}}\sum_{i=1}^{n}I_{\D}(t)^{-1}I(U_i\geqslant t>L_i)\big\{I(X_i\geqslant t>L_i) - S(t)\big\}+o_p(1),
\end{align}
where $\bar{I}_n(t) = \frac{1}{n}\sum_{i=1}^{n}I(U_i\geqslant t>L_i)$, which converges to $I_{\D}(t) = E\left[I(U\geqslant t > L) \right]$ as $n \to \infty$. It then follows by the classical Central Limit Theorem (CLT) that $\sqrt{n}\>\big(\Shat_\sd - S(t)\big)$ converges in distribution to zero-mean Gaussian distribution with asymptotic variance 
$\Sigma_\sd=I_{\D}(t)^{-2}E\big[I(U\geqslant t > L) \big\{I(X\geqslant t>L) - S(t)\big\}\big]^{2}$, which can be estimated by 
$$\Sighat_\sd = \frac{1}{n}\sum_{i=1}^{n}\widebar{I}_{n}(t)^{-2} I(U_i\geqslant t>L_i)^2\big\{I(X_i\geqslant t>L_i)- \Shat_\sd\big\}^2.$$
Furthermore, by using Theorem 8.2 (ULLN) of \citet{pollard1990empirical}, it is easy to prove that $\Sighat_\sd$ is uniformly consistent in $t\in\mathcal{T}$.

\subsubsection{Asymptotic properties of the semi-supervised estimator $\Shat_{\D}(t)$}
\label{Dsspf}

Here we prove the asymptotic properties of the semi-supervised estimator $\Shat_{\D}(t)$. Before that, we first derive the consistency and asymptotic normality of $\widehat{\Beta}_{\D}^{t}$. For consistency, from the information equations shown in the main paper, we have
\begin{align*}
    \mathbf{U}_n\left(\Beta_{\D}^t\right) = &\frac{1}{n}\sum_{i=1}^{n}  \Phibar_{\D i}^{t}I(U_i\geqslant t>L_i)\big\{ I(X_i\geqslant t>L_i)- g(\Beta_{\D}^{t\trans}\Phibar_{\D i}^{t})\big\}\\
    =&E\Big[\Phibar_{\D i}^{t} I\left(U_i \geqslant t >L_i\right)\big\{I\left(X_i \geqslant t>L_i\right)-g(\Beta_{\D}^{t\trans} \Phibar_{\D i}^{t})\big\}\Big]+ o_p(1)\\
    =&\widebar{\mathbf{U}}\left(\Beta_{\D}^{t}\right)+o_p(1).
\end{align*}
Thus, from Conditions C3$-$C6, using the same argument in Lemma~\ref{lem1}, we have
\begin{align*}
    &\sup_{\Beta_{\D}^{t}\in \mathcal{B}_{\epsilon \D}, t \in \mathcal{T}}\Big\|\mathbf{U}_n\left(\Beta_{\D}^t\right) - \widebar{\mathbf{U}}\left(\Beta_{\D}^t\right)\Big\|_2\\
    =&\sup_{\Beta_{\D}^{t}\in \mathcal{B}_{\epsilon \D}, t \in \mathcal{T}}\Big\|\frac{1}{n} \sum_{i=1}^n \Phibar_{\D i}^{t} I\left(U_i \geqslant t>L_i\right) \Big[I\left(X_i \geqslant t>L_i\right)-\big\{g(\widebar{\Beta}_{\D}^{t\trans} \Phibar_{\D i}^{t}) +\dot{g}(\widebar{\Beta}_{\D}^{t\trans} \Phibar_{\D i}^{t})\Phibar_{\D i}^{t\trans}\big(\Beta_{\D}^{t} \\ 
    &- \widebar{\Beta}_{\D}^t\big)\big\}\Big]- E\Big[\Phibar_{\D i}^{t} I\left(U_i \geqslant t>L_i\right)\big[I\left(X_i \geqslant t>L_i\right)-\big\{g(\widebar{\Beta}_{\D}^{t\trans} \Phibar_{\D i}^{t}) +\dot{g}(\widebar{\Beta}_{\D}^{t\trans} \Phibar_{\D i}^{t})\Phibar_{\D i}^{t\trans}\big(\Beta_{\D}^{t} - \\ 
    &\widebar{\Beta}_{\D}^t\big)\big\}\big]\Big]\Big \|_2 + o_p(1)\\
    \leqslant &\sup_{\Beta_{\D}^{t}\in \mathcal{B}_{\epsilon \D}, t \in \mathcal{T}}\Big\|-\frac{1}{n} \sum_{i=1}^n \Phibar_{\D i}^{t} I\left(U_i \geqslant t>L_i\right) \dot{g}(\widebar{\Beta}_{\D}^{t\trans} \Phibar_{\D i}^{t})\Phibar_{\D i}^{t\trans}\big(\Beta_{\D}^{t}-\widebar{\Beta}_{\D}^t\big) \\
    & + E\Big[\Phibar_{\D i}^{t} I\big(U_i \geqslant t>L_i\big)\dot{g}(\widebar{\Beta}_{\D}^{t\trans} \Phibar_{\D i}^{t})\Phibar_{\D i}^{t\trans}\big(\Beta_{\D}^{t}-\widebar{\Beta}_{\D}^t\big)\big\}\Big]\Big \|_2\\
    &+\sup_{\Beta_{\D}^{t}\in \mathcal{B}_{\epsilon \D}, t \in \mathcal{T}}\Big\|\frac{1}{n} \sum_{i=1}^n \Phibar_{\D i}^{t} I\left(U_i \geqslant t>L_i\right)\big\{I\left(X_i \geqslant t>L_i\right)-g(\widebar{\Beta}_{\D}^{t\trans} \Phibar_{\D i}^{t})\big\}\\
    &- E\Big[\Phibar_{\D i}^{t} I\big(U_i \geqslant t>L_i\big)\big\{I\left(X_i \geqslant t>L_i\right)-g(\widebar{\Beta}_{\D}^{t\trans} \Phibar_{\D i}^{t})\big\}\Big]\Big \|_2 + o_p(1)\\
    =& o_p(1).
\end{align*}
From Condition C6, it follows directly from Appendix I of \citet{tian2007model} that there exists a finite $\widebar{\Beta}_{\D}^t$ that solve $\widebar{\mathbf{U}}\left(\Beta_{\D}^t\right)$, and that $\widebar{\Beta}_{\D}^t$ is unique. Hence we have $\widehat{\Beta}_{\D}^t \stackrel{p}{\longrightarrow}\widebar{\Beta}_{\D}^t$, as $n\to \infty$.

For the asymptotic normality of $\widehat{\Beta}_{\D}^t$, noting that $\widehat{\Beta}_{\D}^t \stackrel{p}{\longrightarrow}\widebar{\Beta}_{\D}^t$, employing Theorem 5.21 of \citet{van2000asymptotic} and applying Taylor expansion, we have\\
\begin{align*}
    0 = &\sqrt{n}\>\mathbf{U}_n\big(\bm{\widehat{\beta}}_{\D}^t\big) \\
    =& \frac{1}{\sqrt{n}} \sum_{i=1}^n \Phibar_{\D i}^{t} I\left(U_i \geqslant t>L_i\right) \big\{I\left(X_i \geqslant t>L_i\right)-g(\widebar{\Beta}_{\D}^{t\trans} \Phibar_{\D i}^{t})\big\} \\
     &- \frac{1}{n} \sum_{i=1}^n \Phibar_{\D i}^{t\otimes 2} I\left(U_i \geqslant t>L_i\right)\dot{g}(\widebar{\Beta}_{\D}^{t\trans} \Phibar_{\D i}^{t})\>\sqrt{n}\big(\bm{\widehat{\beta}}_{\D}^{t} - \widebar{\Beta}_{\D}^t\big) + o_p(1).
\end{align*}
Thus, we have
\begin{align}
\label{dbetaasy}
    \sqrt{n}\left(\widehat{\Beta}_{\D}^t - \widebar{\Beta}_{\D}^t\right)
    = \frac{1}{\sqrt{n}} \sum_{i=1}^n \textbf{A}_{\D}^{-1} \Phibar_{\D i}^{t} I\left(U_i \geqslant t>L_i\right)\big\{I\left(X_i \geqslant t>L_i\right)-g(\widebar{\Beta}_{\D}^{t \trans} \Phibar_{\D i}^{t})\big\}+o_p(1),
\end{align}
where $\textbf{A}_{\D}=E\left[\Phibar_{\D}^{t\otimes 2} I\left(U \geqslant t>L\right) \dot{g}(\widebar{\Beta}_{\D}^{t \trans} \Phibar_{\D}^{t})\right]$. It then follows by the classical ${\rm C L T}$ that $\sqrt{n}\big(\widehat{\Beta}_{\D}^t-\widebar{\Beta}_{\D}^t\big)\to N\left(0, \mathbf{J}_{\D}\right)$ in distribution, where $\mathbf{J}_{\D}=\textbf{A}_{\D}^{-1} \textbf{V}_{\D} \textbf{A}_{\D}^{-1}$, and $\textbf{V}_{\D}=E\big[\Phibar_{\D}^{t\otimes 2} I\big(U \geqslant t>L\big)^2\big\{I\big(X \geqslant t>L\big)-g(\widebar{\Beta}_{\D}^{t\trans} \Phibar_{\D}^{t})\big\}^2\big]$, which can be estimated by $\widehat{\mathbf{J}}_{\D}=\widehat{\textbf{A}}_{\D}^{-1}$ $\widehat{\textbf{V}}_{\D} \widehat{\textbf{A}}_{\D}^{-1}$, $\widehat{\textbf{A}}_{\D}=\frac{1}{n} \sum_{i=1}^n \Phibar_{\D_i}^{t\otimes 2} I\big(U_i \geqslant t>L_i\big)\dot{g}(\widehat{\Beta}_{\D}^{t \trans} \Phibar_{\D i}^{t})$, and $\widehat{\textbf{V}}_{\D}=\frac{1}{n} \sum_{i=1}^n \Phibar_{\D i}^{t\otimes 2} I\left(U_i \geqslant t>L_i\right)^2\big\{I\big(X_i \geqslant t>L_i\big)-g(\widehat{\Beta}_{\D}^{t\trans} \Phibar_{\D i}^{t})\big\}^2$.

Next, we derive the asymptotic properties of $\Shat_{\D}(t)$. For consistency, we have
\begin{align*}
\Shat_{\D}(t)& = \frac{\sum_{i= n+1}^{n+N}I(U_i\geqslant t>L_i)g(\widehat{\Beta}_{\D}^{t \trans} \Phibar_{\D i}^{t})}{\sum_{i=n+1}^{n+N}I(U_i\geqslant t > L_i)}\\
 & = \frac{\sum_{i= n+1}^{n+N} I\left(U_i \geqslant t>L_i\right)\big\{g(\widebar{\Beta}_{\D}^{t\trans} \Phibar_{\D i}^{t})+\dot{g}(\widebar{\Beta}_{\D}^{t \trans} \Phibar_{\D i}^{t}) \Phibar_{\D i}^{t\trans}\big(\widehat{\Beta}_{\D}^t-\widebar{\Beta}_{\D}^t\big)\big\}}{\sum_{i=n+1}^{n+N}I(U_i\geqslant t > L_i)} + o_p(1).
\end{align*}   
Note that $\widebar{\Beta}_{\D}^t$ is the unique solution to $\widebar{\mathbf{U}}(\Beta_{\D}^{t})$ and that $\Phibar_{\D}^{t}$ contain 1, we obtain
\begin{align*}
S(t)=\frac{E\left[I(U \geqslant t > L) I(X \geqslant t > L)\right] }{ E\left[I( U \geqslant t>L)\right]} = \frac{E\left[I(U \geqslant t>L) g(\widebar{\Beta}_{\D}^{t\trans} \Phibar_{\D}^{t})\right] }{ E\left[I(U \geqslant t>L)\right]} .
\end{align*}
Then, noting that $\widehat{\Beta}_{\D}^t \stackrel{p}{\longrightarrow}\widebar{\Beta}_{\D}^t$, similar to Lemma~\ref{lem1}, we have
\begin{align*}
&\sup_{ t \in \mathcal{T}}\left|\Shat_{\D}(t)-S(t)\right| \\
\leqslant &\sup_{t \in \mathcal{T}}\left| \frac{\sum_{i= n+1}^{n+N}I(U_i\geqslant t>L_i)g(\widebar{\Beta}_{\D}^{t \trans}\Phibar_{\D i}^{t})}{\sum_{i=n+1}^{n+N}I(U_i\geqslant t>L_i)} -\frac{E\big[I(U \geqslant t>L) g(\widebar{\Beta}_{\D}^{t\trans} \Phibar_{\D}^{t})\big]} {E\left[I(U \geqslant t>L)\right]} \right| \\
& + \sup_{t \in \mathcal{T}}\left|\frac{\sum_{i= n+1}^{n+N}I(U_i\geqslant t>L_i)\dot{g}(\widebar{\Beta}_{\D}^{t \trans} \Phibar_{\D i}^{t}) \Phibar_{\D i}^{t\trans}\left(\widehat{\Beta}_{\D}^t-\widebar{\Beta}_{\D}^t\right)}{\sum_{i=n+1}^{n+N}I(U_i\geqslant t>L_i)}\right| =  o_p(1).
\end{align*}
Hence, $\Shat_{\D}(t)$ is uniformly consistent in $t\in \mathcal{T}$.

For asymptotic normality, applying Taylor expansion, we have
\begin{align}
\label{dsslpsy}
&\sqrt{n}\left(\Shat_{\D}(t)-S(t)\right)\\ \nonumber
=&\sqrt{n}\Bigg[\frac{\sum_{i= n+1}^{n+N}I(U_i\geqslant t>L_i)g(\widebar{\Beta}_{\D}^{t \trans}\Phibar_{\D i}^{t})}{\sum_{i=n+1}^{n+N}I(U_i\geqslant t>L_i)}-\frac{E\big[I(U \geqslant t>L) g(\widebar{\Beta}_{\D}^{t\trans} \Phibar_{\D}^{t})\big]} {E\left[I(U \geqslant t>L)\right]}\\\nonumber
&+\frac{\sum_{i= n+1}^{n+N}I(U_i\geqslant t>L_i)\dot{g}(\widebar{\Beta}_{\D}^{t \trans} \Phibar_{\D i}^{t}) \Phibar_{\D i}^{t\trans}\left(\widehat{\Beta}_{\D}^t-\widebar{\Beta}_{\D}^t\right)}{\sum_{i=n+1}^{n+N}I(U_i\geqslant t>L_i)}+o_p\big(\|\widehat{\Beta}_{\D}^t - \widebar{\Beta}_{\D}^t\|_2\big)\Bigg]\\\nonumber
=&\frac{\sum_{i= n+1}^{n+N} I\left(U_i \geqslant t>L_i\right) \dot{g}(\widebar{\Beta}_{\D}^{t\trans} \bar{\bm\Phi}_{\D i})\bar{\bm\Phi}_{\D i}^{\trans} \sqrt{n}\left(\widehat{\Beta}_{\D}^t - \widebar{\Beta}_{\D}^t\right)}{\sum_{i= n+1}^{n+N} I\left(U_i \geqslant t>L_i\right)}+O_p\big(\sqrt{n}/\sqrt{N}\big)+o_p(1)\\\nonumber
=&I_{\D}(t)^{-1} \textbf{B}_{\D} \sqrt{n}\left(\widehat{\Beta}_{\D}^t - \widebar{\Beta}_{\D}^t\right)+o_p(1),
\end{align}
where $\textbf{B}_{\D} = E\big[I(U\geqslant t>L)\dot{g}(\widebar{\Beta}_{\D}^{t \trans}\Phibar_{\D}^{t}) \Phibar_{\D}^{t\trans}\big]$, which can be estimated by $\widehat{\textbf{B}}_{\D} =\frac{1}{N} \sum_{i=n+1}^{n+N}I\big(U_i$ $\geqslant t>L_i\big)\dot{g}(\widehat{\Beta}_{\D}^{t \trans} \Phibar_{\D i}^{t}) \Phibar_{\D i}^{t\trans}$.
Then the asymptotic normality of $\widehat{\Beta}_{\D}^t$ ensures the weak convergence of $\Shat_{\D}(t)$. Also, \eqref{dbetaasy} and \eqref{dsslpsy} imply that
\begin{align*}
    &\sqrt{n}\left(\Shat_{\D}(t)-S(t)\right)\\
    = &\frac{1}{\sqrt{n}} \sum_{i=1}^n I_{\D}(t)^{-1} \textbf{B}_{\D} \textbf{A}_{\D}^{-1} \Phibar_{\D i}^{t} I\left(U_i \geqslant t>L_i\right)\big\{I\left(X_i \geqslant t>L_i\right)-g(\widebar{\Beta}_{\D}^{t \trans} \Phibar_{\D i}^{t})\big\} + o_p(1).
\end{align*}
Note that 
\begin{align*}
     \textbf{B}_{\D} \textbf{A}_{\D}^{-1} 
    = &E\left[\Phibar_{\D}^{t\trans} I\left(U \geqslant t>L\right)\dot{g}(\widebar{\Beta}_{\D}^{t \trans} \Phibar_{\D}^{t})\right] E\left[ \Phibar_{\D}^{t\otimes 2} I\left(U \geqslant t>L\right)\dot{g}(\widebar{\Beta}_{\D}^{t \trans} \Phibar_{\D}^{t})\right]^{-1}\\
    = & \arg \min_{\bm\alpha}E\left[I\left(U \geqslant t>L\right)\dot{g}(\widebar{\Beta}_{\D}^{t \trans} \Phibar_{\D}^{t})(1- \bm\alpha^{\trans}\Phibar_{\D}^{t})^{2}\right].
\end{align*}
Since $\Phibar_{\D i}^{t}$ contain 1, the minimum $\widehat{\bm \alpha}$ can be chosen such that $1-\bm\alpha^{\trans}\Phibar_{\D i}^{t} = 0$, for $i =1,2, \ldots, (n+N)$, which implies that $\textbf{B}_{\D} \textbf{A}_{\D}^{-1}\Phibar_{\D i}^{t} =1$. Thus 
\begin{align}
\label{dssasy}
    &\sqrt{n}\left(\Shat_{\D}(t)-S(t)\right)\\\nonumber
    = &\frac{1}{\sqrt{n}}\sum_{i=1}^n I_{\D}(t)^{-1} I\left(U_i \geqslant t>L_i\right)\big\{I\left(X_i \geqslant t>L_i\right)-g(\widebar{\Beta}_{\D}^{t \trans} \Phibar_{\D i}^{t})\big\}+o_p(1).
\end{align}
It then follows from the classical CLT that
$$ \sqrt{n}\left(\Shat_{\D}(t)-S(t)\right)\stackrel{d}{\longrightarrow} N\left(0, \Sigma_{\D}(t) \right), $$
where 
$\Sigma_{\D}(t)=I_{\D}(t)^{-2}E\big[ I\left(U \geqslant t > L \right)\big\{I\left(X \geqslant t>L\right)-g(\widebar{\Beta}_{\D}^{t \trans} \Phibar_{\D}^{t})\big\}\big]^{2}$, which can be estimated by
$$\Sighat_{\D}(t) = \frac{\frac{1}{n} \sum_{i=1}^n I\left(U_i \geqslant t>L_i\right)^2\big\{I\left(X_i \geqslant t>L_i\right)-g(\widehat{\Beta}_{\D}^{t \trans} \Phibar_{\D i}^{t})\big\}^{2}}{\big\{\frac{1}{N}\sum_{i=n+1}^{n+N}I(U_i\geqslant t > L_i)\big\}^{2}}.$$
Similar to Lemma~\ref{lem1}, it is easy to prove that $\Sighat_{\D}(t)$ is uniformly consistent in $t\in\mathcal{T}$.

\subsubsection{Asymptotic properties of the intrinsic semi-supervised estimator $\Shat_\ssd$}
\label{DIsspf}

Now we derive the asymptotic properties of the intrinsic semi-supervised estimator $\Shat_\ssd$. Similar to the previous section, we first derive the consistency and asymptotic normality of $\widehat{\Beta}_{\ID}^t$. For consistency, we can adapt the same procedure as in Lemma~\ref{lem1}. We obtain
\begin{align*}
    &\sup_{\Beta_{\D}^{t} \in \mathcal{B}_{\epsilon \D}, t\in\mathcal{T}}\Big\|\frac{1}{n} \sum_{i=1}^n \bar{I}_{N}(t)^{-2} I\left(U_i \geqslant t>L_i\right)^2\big\{I\left(X_i \geqslant t>L_i\right)-g(\Beta_{\D}^{t \trans} \Phibar_{\D i}^{t})\big\}^{2}\\
    &- E\Big[I\left(U \geqslant t > L\right)^2\big\{I\left(X \geqslant t > L\right)-g(\Beta_{\D}^{t \trans} \Phibar_{\D}^{t})\big\}^2 \Big] I_{\D}(t)^{-2} \Big\|_2 = o_p(1),
\end{align*}
where $\bar{I}_{N}(t) = \frac{1}{N}\sum_{i=n+1}^{n+N}I(U_i\geqslant t > L_i)$.
Also, 
\begin{align*}
    &\sup_{\Beta_{\D}^{t} \in \mathcal{B}_{\epsilon \D}, t\in\mathcal{T}}\Big\|\frac{1}{n} \sum_{i=1}^n I\left(U_i \geqslant t>L_i\right) \big\{I\left(X_i \geqslant t>L_i\right)-g(\Beta_{\D}^{t \trans} \Phibar_{\D i}^{t})\big\}\\
    & - E\Big[I\left(U \geqslant t > L\right)\big\{I\left(X \geqslant t > L \right)-g(\Beta_{\D}^{t \trans} \Phibar_{\D}^{t})\big\}\Big] \Big\|_{2} = o_p(1).
\end{align*}
From Condition C6, it follows directly from Appendix I of \citet{tian2007model} that there exists a finite $\widebar{\Beta}_{\ID}^t$ that satisfy \eqref{dintrexp} in the main paper, and that $\widebar{\Beta}_{\ID}^t$ is unique. This implies $\widehat{\Beta}_{\ID}^{t} \stackrel{p}{\longrightarrow} \widebar{\Beta}_{\ID}^{t}$, as $n\to \infty$.

For asymptotic normality, from the first equation of \eqref{dintreq} given in the main paper, denoted as $\textbf{I}_{(4)}$, we apply second-order Taylor expansion at $\widebar{\Beta}_{\ID}^{t}$,
\begin{align*}
\textbf{I}_{(4)}=&\frac{1}{n} \sum_{i=1}^n \bar{I}_{N}(t)^{-2} I\left(U_i \geqslant t>L_i\right)^2\big\{I\left(X_i \geqslant t>L_i\right)-g(\widebar{\Beta}_{\ID}^{t \trans} \Phibar_{\D i}^{t})\big\}^2\\
 &+ \left(\Beta_{\ID}^t-\widebar{\Beta}_{\ID}^t\right)^{\trans} \Big[\frac{1}{n} \sum_{i=1}^n \bar{I}_{N}(t)^{-2} I\left(U_i \geqslant t>L_i\right)^2 \big[ \dot{g}^{2}(\widebar{\Beta}_{\ID}^{t\trans} \Phibar_{\D i}^{t}) - \big\{I\big(X_i\geqslant t>L_i\big)\\
&-g(\widebar{\Beta}_{\ID}^{t\trans} \Phibar_{\D i}^{t})\big\} \ddot{g}(\widebar{\Beta}_{\ID}^{t\trans} \Phibar_{\D i}^{t})\big]\Phibar_{\D i}^{t\otimes 2}\left(\Beta_{\ID}^t-\widebar{\Beta}_{\ID}^t\right)- \frac{2}{n}\sum_{i=1}^n \bar{I}_{N}(t)^{-2} I\big(U_i \geqslant t>L_i\big)^2\\
&\big\{I\left(X_i \geqslant t>L_i\right)-g(\widebar{\Beta}_{\ID}^{t\trans} \Phibar_{\D i}^{t})\big\} \dot{g}(\widebar{\Beta}_{\ID}^{t\trans} \Phibar_{\D i}^{t})\Phibar_{\D i}^{t} + o_p\big(\|\widehat{\Beta}_{\ID}^t- \widebar{\Beta}_{\ID}^t\|_2 + n^{-\frac{1}{2}}\big)\Big] .
\end{align*}
From the second equation of \eqref{dintreq} in the main paper, denoted as $\textbf{II}_{(4)}$, we apply first-order Taylor expansion at $\widebar{\Beta}_{\ID}^{t}$,
\begin{align*}
    \textbf{II}_{(4)} = & \frac{1}{n}\sum_{i=1}^n I\left(U_i \geqslant t>L_i\right)\big\{I\left(X_i \geqslant t>L_i\right)-g(\widebar{\Beta}_{\ID}^{t\trans} \Phibar_{\D i}^{t})\big\} - \frac{1}{n}\sum_{i=1}^n I\big(U_i \geqslant t>L_i\big)\\
    &\dot{g}(\widebar{\Beta}_{\ID}^{t\trans} \Phibar_{\D i}^{t})\Phibar_{\D i}^{t\trans} \big(\Beta_{\ID}^t-\widebar{\Beta}_{\ID}^t\big) + o_p\big(\|\widehat{\Beta}_{\ID}^t-\widebar{\Beta}_{\ID}^t\|_2 +n^{-\frac{1}{2}} \big) = 0.
\end{align*}
Then, we have
\begin{align*}
    &\widehat{\Beta}_{\ID}^t= \arg\min\limits_{\Beta_{\D}^t} \left(\Beta_{\ID}^t-\widebar{\Beta}_{\ID}^t\right)^{\trans}\Big[-\textbf{A}_1 \left(\Beta_{\ID}^t-\widebar{\Beta}_{\ID}^t\right)-\bm{\Sigma}_{11} + o_p\big(\| \widehat{\Beta}_{\ID}^t-\widebar{\Beta}_{\ID}^t\|_2 + n^{-\frac{1}{2}}\big)\Big],\\
    & \qquad \qquad  \quad \text{s.t.}  \quad \textbf{B}_1^{\trans}\big(\Beta_{\ID}^t-\widebar{\Beta}_{\ID}^t\big) - \bm{\Sigma}_{12} - o_p\big(\|\widehat{\Beta}_{\ID}^t-\widebar{\Beta}_{\ID}^t\|_2 + n^{-\frac{1}{2}}\big) =0,
\end{align*}
where $\textbf{A}_1$ and $\textbf{B}_1$ are defined as above, and 
\begin{align*}
    \bm{\Sigma}_{11} = &\frac{1}{n}\sum_{i=1}^n \bar{I}_{N}(t)^{-2} I\left(U_i \geqslant t>L_i\right)^2\big\{I\big(X_i \geqslant t>L_i\big)-g(\widebar{\Beta}_{\ID}^{t\trans} \Phibar_{\D i}^{t})\big\} \dot{g}(\widebar{\Beta}_{\ID}^{t\trans}\Phibar_{\D i}^{t})\Phibar_{\D i}^{t};\\
    \bm{\Sigma}_{12} = &\frac{1}{n}\sum_{i=1}^n I\big(U_i \geqslant t>L_i\big)\big\{I\big(X_i \geqslant t>L_i\big)-g(\widebar{\Beta}_{\ID}^{t\trans} \Phibar_{\D i}^{t})\big\}.
\end{align*}
Then from Theorem 5.21 of \citet{van2000asymptotic}, we obtain
\begin{align*}
    \widehat{\Beta}_{\ID}^t-\widebar{\Beta}_{\ID}^t 
    = & \big[-\textbf{A}_1^{-1} + \textbf{A}_1^{-1}\textbf{B}_1(\textbf{B}_1^{\trans}\textbf{A}_1^{-1}\textbf{B}_1)^{-1}\textbf{B}_1^{\trans}\textbf{A}_1^{-1} \big]\bm{\Sigma}_{11} +\textbf{A}_1^{-1}\textbf{B}_1(\textbf{B}_1^{\trans}\textbf{A}_1^{-1}\textbf{B}_1)^{-1}\bm{\Sigma}_{12} \\
    = & O_p(n^{-\frac{1}{2}}).
\end{align*}
It then follows from the classical CLT that $\sqrt{n}\> \big( \widehat{\Beta}_{\ID}^t-\widebar{\Beta}_{\ID}^t \big)$ is asymptotically normal. Hence we can use the same argument of the proof for the semi-supervised estimator $\Shat_{\D}(t)$ to show that $\sup_{t\in\mathcal{T}}\left|\Shat_\ssd-S(t)\right| = o_p(1)$. From \eqref{dintrexp} given in the main paper, we have
 $$S(t)=\frac{E\left[I(U \geqslant t > L) I(X \geqslant t > L)\right] }{ E\left[I( U \geqslant t>L)\right]} = \frac{E\left[I(U \geqslant t>L) g(\widebar{\Beta}_{\ID}^{t\trans} \bar{\bm\Phi}_{\D})\right] }{ E\left[I(U \geqslant t>L)\right]} .$$
Then
\begin{align}
\label{dintrpsy}
&\sqrt{n}\left(\Shat_\ssd-S(t)\right)\\ \nonumber
=&\sqrt{n}\Bigg[\frac{\sum_{i= n+1}^{n+N}I(U_i\geqslant t>L_i)g(\widebar{\Beta}_{\ID}^{t \trans}{\bar{\bm\Phi}}_{\D i})}{\sum_{i=n+1}^{n+N}I(U_i\geqslant t>L_i)}-\frac{E\big[I(U \geqslant t>L) g(\widebar{\Beta}_{\ID}^{t\trans} \bar{\bm\Phi}_{\D})\big]} {E\left[I(U \geqslant t>L)\right]}\\\nonumber
&+\frac{\sum_{i= n+1}^{n+N}I(U_i\geqslant t>L_i)\dot{g}(\widebar{\Beta}_{\ID}^{t \trans} \Phibar_{\D i}^{t}) \Phibar_{\D i}^{t\trans}\left(\widehat{\Beta}_{\ID}^t-\widebar{\Beta}_{\ID}^t\right)}{\sum_{i=n+1}^{n+N}I(U_i\geqslant t>L_i)}+o_p\big(\|\widehat{\Beta}_{\ID}^t-\widebar{\Beta}_{\ID}^t\|_2\big)\Bigg]\\\nonumber
=&\frac{\sum_{i= n+1}^{n+N} I\left(U_i \geqslant t>L_i\right) \dot{g}(\widebar{\Beta}_{\ID}^{t\trans} \bar{\bm\Phi}_{\D i})\bar{\bm\Phi}_{\D i}^{\trans} \sqrt{n}\left(\widehat{\Beta}_{\ID}^t - \widebar{\Beta}_{\ID}^t\right)}{\sum_{i= n+1}^{n+N} I\left(U_i \geqslant t>L_i\right)}+O_p\big(\sqrt{n}/\sqrt{N}\big)+o_p(1)\\\nonumber
=&I_{\D}(t)^{-1}\sqrt{n} \textbf{B}_{1}^{\trans}\left(\widehat{\Beta}_{\ID}^t - \widebar{\Beta}_{\ID}^t\right)+o_p(1)\\\nonumber
= & I_{\D}(t)^{-1}\sqrt{n} \bm{\Sigma}_{12} + o_p(1)\\\nonumber
= &\frac{1}{\sqrt{n}} \sum_{i=1}^n I_{\D}(t)^{-1} I\left(U_i \geqslant t>L_i\right)\big\{I\left(X_i \geqslant t>L_i\right)-g(\widebar{\Beta}_{\ID}^{t \trans} \Phibar_{\D i}^{t})\big\}+o_p(1).
\end{align}
It then follows from the classical CLT that
$$ \sqrt{n}\left(\Shat_\ssd-S(t)\right)\stackrel{d}{\longrightarrow} N\left(0, \Sigma_\ssd\right), $$
where
$\Sigma_\ssd=I_{\D}(t)^{-2}E\left[ I\left(U \geqslant t>L\right)\big\{I\left(X \geqslant t>L\right)-g(\widebar{\Beta}_{\ID}^{t \trans} \Phibar_{\D}^{t})\big\}\right]^{2}$, which has a uniformly consistent estimator
$$\Sighat_\ssd = \frac{\frac{1}{n} \sum_{i=1}^n I\left(U_i \geqslant t>L_i\right)^2\big\{I\left(X_i \geqslant t>L_i\right)-g(\widehat{\Beta}_{\ID}^{t \trans} \Phibar_{\D i}^{t})\big\}^{2}}{\big\{\frac{1}{N}\sum_{i=n+1}^{n+N}I(U_i\geqslant t > L_i)\big\}^{2}}.$$
We then see that
\begin{align*}
&\Sigma_\sd -\Sigma_\ssd \\
= &I_{\D}(t)^{-2}E\big[I(U\geqslant t > L) \big\{I(X\geqslant t>L) - S(t)\big\}\big]^{2}\\
& - I_{\D}(t)^{-2}E\big[ I(U \geqslant t > L)\big\{I(X \geqslant t>L)-g(\widebar{\Beta}_{\ID}^{t \trans} \Phibar_{\D}^{t})\big\}\big]^{2}\\
=& I_{\D}(t)^{-2}E\Big[ I(U \geqslant t > L)\big\{g(\widebar{\Beta}_{\ID}^{t \trans} \Phibar_{\D}^{t})-S(t)\big\}^2 \\
&+ 2 I(U \geqslant t > L)\big\{I(X \geqslant t>L)-g(\widebar{\Beta}_{\ID}^{t \trans} \Phibar_{\D}^{t})\big\}\big\{g(\widebar{\Beta}_{\ID}^{t \trans} \Phibar_{\D}^{t})-S(t)\big\}\Big].
\end{align*}
Therefore, when the imputation model \eqref{dmodel} given in the main paper is correctly specified, it follows that 
$$\Sigma_\sd -\Sigma_\ssd =  I_{\D}(t)^{-2}E\Big[ I(U \geqslant t > L)\big\{g(\widebar{\Beta}_{\ID}^{t \trans} \Phibar_{\D}^{t})-S(t)\big\}\Big]^2 \succeq \bm{0}.$$

\subsection{Asymptotic properties of the estimators based on the left (right) censoring label}
\label{Asymp}

In this section, we prove the asymptotic properties of the supervised estimator $\Shat_\sl$, semi-supervised estimator $\Shat_{\L}$, intrinsic semi-supervised estimator $\Shat_\ssl$, respectively, based on the left  censoring label. The asymptotic properties of the estimators based on the right censoring label can be proved similarly.

\subsubsection{Asymptotic properties of the supervised estimator $\Shat_\sl$}
\label{Lslpf}
We derive the asymptotic properties of $\Shat_\sl$ as follows. For consistency,
\begin{align*}
    \Shat_\sl& = \sum_{i=1}^{n}K_{h_l}(L_i - t) I(T_i \geqslant L_i) \big/ \sum_{i=1}^{n}K_{h_l}(L_i - t)\\
    & = \int K(u)\diff u f_l(t) E\left[I(T \geqslant t)\right] \big/ \big\{f_l(t) \int K(u)\diff u\big\}+ o_p(1)\\
    & = S(t) + o_p(1),
\end{align*}
where $K(\cdot)$ is the kernel function, and $f_l(t)$ is the density function of $L$. Then under the above mild regularity conditions, from Theorem 8.2 (ULLN) of \citet{pollard1990empirical}, we have $\Shat_\sl \stackrel{p}{\longrightarrow} S(t)$ uniformly in $t\in\mathcal{T}$.

Next, we consider the asymptotic normality,
\begin{align}
\label{lsupasy}
    \sqrt{nh_l}\>\left(\Shat_\sl - S(t)\right) 
    = &\sqrt{\frac{h_l}{n}}\sum_{i=1}^{n}\widebar{K}_{nl}(t)^{-1} K_{h_l}(L_i - t)\big\{I(T_i\geqslant L_i) - S(t)\big\}\\ \nonumber
    = & \sqrt{\frac{h_l}{n}}\sum_{i=1}^{n}f_l(t)^{-1}K_{h_l}(L_i - t)\big\{I(T_i\geqslant L_i) - S(t)\big\}+o_p(1),
\end{align}
where $\widebar{K}_{nl}(t) = \frac{1}{n}\sum_{i=1}^{n}K_{h_{l}}(L_i - t)$, which goes to $f_l(t)$ as $h_l\to 0$, $n\to \infty$, and $nh_l\to \infty$.
It then follows by the classical CLT that $\sqrt{nh_l}\>\big(\Shat_\sl - S(t)\big)$ converges in distribution to zero-mean Gaussian distribution with asymptotic variance 
$\Sigma_\sl=\nu_2 f_l(t)^{-1}E\big[I(T\geqslant t ) - S(t)\big]^{2}$, which has a uniformly consistent estimator
$$\Sighat_\sl=\frac{h_{l}}{n}\sum_{i=1}^{n} \widebar{K}_{nl}(t)^{-2} K_{h_l}^{2}(L_i -t)\big\{I(T_i \geqslant L_i) - \Shat(t)\big\}^2.$$

\subsubsection{Asymptotic properties of the semi-supervised estimator $\Shat_{\L}(t)$}
\label{Lsspf}

To prove the asymptotic properties of the semi-supervised estimator $\Shat_{\L}(t)$, we first derive the consistency and asymptotic normality of $\widehat{\Beta}_{\L}^{t}$. For consistency,
\begin{align*}
    \mathbf{U}_n(\Beta_{\L}^t)&=\frac{1}{n}\sum_{i=1}^{n}K_{h_l}(L_i-t)\Phibar_{\L i}^{t}\big\{I(T_i\geqslant L_i)-g(\Beta_{\L}^{t\trans}\Phibar_{\L i}^{t})\big\}\\
    &=E\big[K_{h_l}(L_i-t)\Phibar_{\L i}^{t} \big\{I(T_i \geqslant L_i)-g(\Beta_{\L}^{t \trans}\Phibar_{\L i}^{t})\big\}\big] + o_p(1)\\
    &=\widebar{\mathbf{U}}(\Beta_{\L}^t) + o_p(1).
\end{align*}
From Lemma~\ref{lem1}, we can obtain $\sup_{\Beta_{\L}^t \in \mathcal{B}_{\epsilon \L}, t\in\mathcal{T}}\big\| \mathbf{U}_n(\Beta_{\L}^t) - \widebar{\mathbf{U}}(\Beta_{\L}^t) \big\|_2 = o_p(1)$. From Condition C6, it follows directly from Appendix I of \citet{tian2007model} that there exists a finite $\widebar{\Beta}_{\L}^t$ that solve $\widebar{\mathbf{U}}\left(\Beta_{\L}^t\right)$, and that $\widebar{\Beta}_{\L}^t$ is unique. Hence we have $\widehat{\Beta}_{\L}^t \stackrel{p}{\longrightarrow}\widebar{\Beta}_{\L}^t$, as $n\to \infty$.

From the information equations shown in the main paper, Noting that $\widehat{\Beta}_{\L}^t \stackrel{p}{\longrightarrow}\widebar{\Beta}_{\L}^t$, similar to Section~\ref{Dsspf}, applying Theorem 5.21 \citet{van2000asymptotic}, we have
\begin{align*}
    0=&\sqrt{n h_l}\>\mathbf{U}_n(\widehat{\Beta}_{\L}^t)\\
    =&\sqrt{\frac{h_l}{n}}\sum_{i=1}^{n}K_{h_l}(L_i-t)\Phibar_{\L i}^{t}\big\{I(T_i\geqslant L_i)-g(\widebar{\Beta}_{\L}^{t\trans} \Phibar_{\L i}^{t})\big\}\\
    & - \frac{1}{n}\sum_{i=1}^{n}K_{h_l}(L_i-t)\Phibar_{\L i}^{t\otimes 2}\dot{g}(\widebar{\Beta}_{\L}^{t\trans} \Phibar_{\L i}^{t})\sqrt{nh_l}\big(\bm{\widehat{\beta}}_{\L}^{t} - \widebar{\Beta}_{\L}^t\big) + o_p(1).
\end{align*}
Thus, we derive the asymptotic expansion of $\sqrt{nh_{l}}\big(\widehat{\Beta}_{\L}^t-\widebar{\Beta}_{\L}^t\big)$ as follows,
\begin{align}
\label{lbetaasy}
\sqrt{nh_{l}}\left(\widehat{\Beta}_{\L}^t-\widebar{\Beta}_{\L}^t\right)
= \sqrt{\frac{h_{l}}{n}} \sum_{i=1}^{n} \textbf{A}_{\L}^{-1} \Phibar_{\L i}^{t} K_{h_l}(L_i-t) \big\{ I(T_i\geqslant L_i)- g(\widebar{\Beta}_{\L}^{t\trans}\Phibar_{\L i}^{t})\big\} + o_p(1),
\end{align}
where $\textbf{A}_{\L}=f_l(t) E\big[\Phibar_{\L}^{t\otimes 2}\> \dot{g}(\widebar{\Beta}_{\L}^{t\trans}\Phibar_{\L}^{t})\big]$. By using CLT, we have $\sqrt{nh_{l}}\big(\widehat{\Beta}_{\L}^t-\widebar{\Beta}_{\L}^t\big)\to N\left(0, \mathbf{J}_{\L} \right)$ in distribution, where $\mathbf{J}_{\L}=\textbf{A}_{\L}^{-1} \textbf{V}_{\L} \textbf{A}_{\L}^{-1}$, and $\textbf{V}_{\L}=\nu_2 f_l(t) E\big[\Phibar_{\L}^{t\otimes 2}\big\{ I(T \geqslant t)- g(\widebar{\Beta}_{\L}^{t\trans}\Phibar_{\L}^{t})\}^2\big]$. It is easy to obtain a consistent estimator for $\mathbf{J}_{\L}$ as $\widehat{\mathbf{J}}_{\L} = \widehat{\textbf{A}}_{\L}^{-1} \widehat{\textbf{V}}_{\L} \widehat{\textbf{A}}_{\L}^{-1}$, where
$\widehat{\textbf{A}}_{\L} = \frac{1}{n}\sum_{i=1}^{n}K_{h_l}(L_i -t)\Phibar_{\L i}^{t\otimes 2}\> \dot{g}(\widehat{\Beta}_{\L}^{t\trans}\Phibar_{\L i}^{t})$, and $\widehat{\textbf{V}}_{\L} =\frac{h_l}{n} \sum_{i=1}^{n}\Phibar_{\L i}^{t\otimes 2} K_{h_l}^{2}(L_i-t) \big\{ I(T_i\geqslant L_i)- g(\widehat{\Beta}_{\L}^{t\trans}\Phibar_{\L i}^{t})\big\}^2$.

Now we derive the asymptotic properties of $\Shat_{\L}(t)$. For consistency, from \eqref{lmodel} in the main paper and note that $\widehat{\Beta}_{\L}^t \stackrel{p}{\longrightarrow}\widebar{\Beta}_{\L}^t$, applying the Delta method, we have 
\begin{align*}
    &\sup_{t\in\mathcal{T}}\left |\Shat_{\L}(t) - S(t)\right | \\
    = &\sup_{t\in\mathcal{T}}\left|\frac{1}{N}\sum_{i=n+1}^{n+N}\widebar{K}_{\NL}(t)^{-1}K_{h_{\L}}(L_i -t)g(\widehat{\Beta}_{\L}^{t\trans}\Phibar_{\L i}^{t}) - f_l(t)^{-1}E\bigl[K_{h_{\L}}(L-t)g(\widebar{\Beta}_{\L}^{t\trans}\Phibar_{\L}^{t})\bigr]\right| + o_p(1)\\
    \leqslant &\sup_{t\in\mathcal{T}}\left|\frac{1}{N}\sum_{i=n+1}^{n+N}\widebar{K}_{\NL}(t)^{-1}K_{h_{\L}}(L_i -t)g(\widebar{\Beta}_{\L}^{t\trans}\Phibar_{\L i}^{t}) - f_l(t)^{-1}E\bigl[K_{h_{\L}}(L-t)g(\widebar{\Beta}_{\L}^{t\trans}\Phibar_{\L}^{t})\bigr]\right|\\
    & + \sup_{t\in\mathcal{T}}\left|\frac{1}{N}\sum_{i=n+1}^{n+N}\widebar{K}_{\NL}(t)^{-1}K_{h_{\L}}(L_i -t)\dot{g}(\widebar{\Beta}_{\L}^{t\trans}\Phibar_{\L i}^{t})\Phibar_{\L i}^{t\trans}\left(\widehat{\Beta}_{\L}^{t} - \widebar{\Beta}_{\L}^{t}\right) \right| + o_p(1)\\
    = & o_p(1),
    \end{align*}
    where $\widebar{K}_{\NL}(t) = \frac{1}{N}\sum_{i=n+1}^{n+N}K_{h_{\L}}(L_i - t)$, which goes to $f_l(t)$ as $h_{\L}\to 0$, $N\to \infty$, and $Nh_{\L}\to \infty$. Thus, $\Shat_{\L}(t)$ is uniformly consistent in $t\in\mathcal{T}$.

For asymptotic normality, applying Taylor expansion, we have
\begin{align}
\label{lsslpsy}
    &\sqrt{nh_l}\left( \Shat_{\L}(t) - S(t) \right) \\\nonumber
    =&\sqrt{nh_l} \Big[\frac{1}{N}\sum_{i=n+1}^{n+N}\widebar{K}_{\NL}(t)^{-1}K_{h_{\L}}(L_i -t)g(\widehat{\Beta}_{\L}^{t\trans}\Phibar_{\L i}^{t}) - \frac{E\big[K_{h_{\L}}(L-t)g(\widebar{\Beta}_{\L}^{t\trans}\Phibar_{\L}^{t})\big]}{E\left[K_{h_{\L}}(L-t) \right]} \\\nonumber
    &+\frac{E\big[K_{h_{\L}}(L-t)g(\widebar{\Beta}_{\L}^{t\trans}\Phibar_{\L}^{t})\big]}{E\left[K_{h_{\L}}(L-t) \right]} - S(t)\Big] \\\nonumber
    = & \sqrt{nh_l}\Big[\frac{1}{N}\sum_{i=n+1}^{n+N}\widebar{K}_{\NL}(t)^{-1}K_{h_{\L}}(L_i -t)g(\widebar{\Beta}_{\L}^{t\trans}\Phibar_{\L i}^{t})-\frac{E\big[K_{h_{\L}}(L-t)g(\widebar{\Beta}_{\L}^{t\trans}\Phibar_{\L}^{t})\big]}{E\left[K_{h_{\L}}(L-t) \right]}\\\nonumber
    &+\frac{1}{N}\sum_{i=n+1}^{n+N}\widebar{K}_{\NL}(t)^{-1}K_{h_{\L}}(L_i -t)\dot{g}(\widebar{\Beta}_{\L}^{t\trans}\Phibar_{\L i}^{t}) \Phibar_{\L i}^{t\trans}(\widehat{\Beta}_{\L}^{t}-\widebar{\Beta}_{\L}^{t}) +o_p\big(\|\widehat{\Beta}_{\L}^{t}-\widebar{\Beta}_{\L}^{t}\|_2\big) \Big]\\\nonumber
    =& \frac{1}{N}\sum_{i=n+1}^{n+N}f_l(t)^{-1}K_{h_{\L}}(L_i -t)\dot{g}(\widebar{\Beta}_{\L}^{t\trans}\Phibar_{\L i}^{t}) \Phibar_{\L i}^{t\trans}\sqrt{nh_l}(\widehat{\Beta}_{\L}^{t}-\widebar{\Beta}_{\L}^{t}) + O_p\big(\sqrt{nh_l}/\sqrt{Nh_{\L}}\big) + o_p(1)\\\nonumber
    =& f_l(t)^{-1} \textbf{B}_{\L} \sqrt{nh_l}\left(\widehat{\Beta}_{\L}^{t}-\widebar{\Beta}_{\L}^{t}\right) + o_p(1),
\end{align}
where
$\textbf{B}_{\L} = f_l(t) E\big[\Phibar_{\L}^{t\trans}  \dot{g}(\widebar{\Beta}_{\L}^{t\trans}\Phibar_{\L}^{t})\big]$, which can be estimated by $\widehat{\textbf{B}}_{\L} = \frac{1}{N}\sum_{i=n+1}^{n+N} K_{h_{\L}}(L_i -t)\dot{g}(\widehat{\Beta}_{\L}^{t\trans}\Phibar_{\L i}^{t}) \Phibar_{\L i}^{t\trans}$.
Then the asymptotic normality of $\widehat{\Beta}_{\L}^t$ ensures the weak convergence of $\Shat_{\L}(t)$. Also, \eqref{lbetaasy} and \eqref{lsslpsy} imply that
\begin{align*}
    &\sqrt{nh_l}\left( \Shat_{\L}(t) - S(t) \right) \\ \nonumber
     =& \sqrt{\frac{h_{l}}{n}}\sum_{i=1}^{n} f_l(t)^{-1} \textbf{B}_{\L} \textbf{A}_{\L}^{-1} \Phibar_{\L i}^{t} K_{h_l}(L_i-t) \big\{ I(T_i\geqslant L_i)- g(\widebar{\Beta}_{\L}^{t\trans}\Phibar_{\L i}^{t})\big\} + o_p(1).
\end{align*}
Then using the same argument to Section~\ref{Dsspf}, we have
\begin{align}
\label{lssasy}
    \sqrt{nh_l}\left( \Shat_{\L}(t) - S(t) \right)=\sqrt{\frac{h_{l}}{n}} \sum_{i=1}^{n}f_l(t)^{-1} K_{h_l}(L_i-t) \big\{ I(T_i\geqslant L_i)- g(\widebar{\Beta}_{\L}^{t\trans}\Phibar_{\L i}^{t})\big\} + o_p(1).
\end{align}
It then follows the classical CLT that
$$\sqrt{nh_l}\left( \Shat_{\L}(t) - S(t) \right) \stackrel{d}{\longrightarrow} N\left(0, \Sigma_{\L}(t)\right), $$
where $\Sigma_{\L}(t)=\nu_2 f_l(t)^{-1}E\big[I(T\geqslant t) - g(\widebar{\Beta}_{\L}^{t\trans}\Phibar_{\L}^{t})\big]^2$, which has the following uniformly consistent estimator
\begin{equation*}
    \Sighat_{\L}(t)=\frac{h_{l}}{n}\sum_{i=1}^{n} \widebar{K}_{\NL}(t)^{-2} K_{h_l}^2(L_i-t)\big\{I(T_i\geqslant L_i) - g(\bm{\widehat{\beta}}_{\L}^{t \trans}\Phibar_{\L i}^{t})\big\}^2.
\end{equation*}

\subsubsection{Asymptotic properties of the intrinsic semi-supervised estimator $\Shat_\ssl$}
\label{LIsspf}

To derive the asymptotic properties for the intrinsic semi-supervised estimator $\Shat_\ssl$, we first derive the consistency and asymptotic normality of $\widehat{\Beta}_{\IL}^t$. For consistency, using the same argument of the proof for Lemma~\ref{lem1}, we obtain
\begin{align*}
    & \sup_{\Beta_{\L}^{t} \in \mathcal{B}_{\epsilon \L}, t\in \mathcal{T}}\Big\|\frac{h_{l}}{n}\sum_{i=1}^{n} \widebar{K}_{\NL}(t)^{-2} K_{h_l}^2(L_i-t)\big\{I(T_i\geqslant L_i) - g(\Beta_{\L}^{t \trans}\Phibar_{\L i}^{t})\big\}^2 \\
    & -\nu_2 f_{l}(t)^{-1} E\big[I(T \geqslant t)-g(\Beta_{\L}^{t \trans}\Phibar_{\L}^{t})\big]^{2} \Big\|_2 = o_p(1),
\end{align*}
and
\begin{align*}
    &\sup_{\Beta_{\L}^{t} \in \mathcal{B}_{\epsilon \L}, t\in \mathcal{T}}\Big\|\frac{1}{n}\sum_{i=1}^{n} K_{h_l}(L_i-t)\big\{I(T_i\geqslant L_i)-g(\Beta_{\L}^{t \trans}\Phibar_{\L i}^{t})\big\} \\
    & -f_{l}(t) E\big[ I(T \geqslant t)- g(\Beta_{\L}^{t \trans}\Phibar_L)\big] \Big\|_{2} = o_p(1).
\end{align*}
From Condition C6, it follows directly from Appendix I of \citet{tian2007model} that there exists a finite $\widebar{\Beta}_{\IL}^t$ that satisfy \eqref{Lintrexp} in the main paper, and that $\widebar{\Beta}_{\IL}^t$ is unique.This implies $\widehat{\Beta}_{\IL}^{t} \stackrel{p}{\longrightarrow} \widebar{\Beta}_{\IL}^{t}.$ 

For asymptotic normality, from the first equation of \eqref{Lintreq} in the main paper, denoted as $\textbf{I}_{(13)}$, we apply second-order Taylor expansion at $\widebar{\Beta}_{\IL}^{t}$ as follows,
\begin{align*}
\textbf{I}_{(13)}=&\frac{h_l}{n} \sum_{i=1}^n \widebar{K}_{\NL}(t)^{-2} K_{h_l}^2(L_i -t)\big\{I\left(T_i \geqslant L_i\right)-g(\widebar{\Beta}_{\IL}^{t \trans} \Phibar_{\L i}^{t})\big\}^2\\
 & +\left(\Beta_{\IL}^t-\widebar{\Beta}_{\IL}^t\right)^{\trans} \Big[\frac{h_l}{n} \sum_{i=1}^n \widebar{K}_{\NL}(t)^{-2} K_{h_l}^2(L_i -t) \big[ \dot{g}^{2}(\widebar{\Beta}_{\IL}^{t\trans} \Phibar_{\L i}^{t}) - \big\{I\left(T_i \geqslant L_i\right)-g(\widebar{\Beta}_{\IL}^{t \trans} \Phibar_{\L i}^{t})\big\}\\
&\ddot{g}(\widebar{\Beta}_{\IL}^{t\trans} \Phibar_{\L i}^{t})\big]\Phibar_{\L i}^{t\otimes 2}\left(\Beta_{\IL}^t-\widebar{\Beta}_{\IL}^t\right)-\frac{2h_l}{n}\sum_{i=1}^n \widebar{K}_{\NL}(t)^{-2} K_{h_l}^2(L_i - t)\big\{I\left(T_i \geqslant L_i\right)-g(\widebar{\Beta}_{\IL}^{t \trans} \Phibar_{\L i}^{t})\big\}\\
&\dot{g}(\widebar{\Beta}_{\IL}^{t\trans} \Phibar_{\L i}^{t})\Phibar_{\L i}^{t} + o_p\big(\|\widehat{\Beta}_{\IL}^t-\widebar{\Beta}_{\IL}^t\|_2 + (nh)^{-\frac{1}{2}}\big)\Big].
\end{align*}
From the second equation of \eqref{Lintreq} in the main paper, denoted as $\textbf{II}_{(13)}$, we apply first-order Taylor expansion at $\widebar{\Beta}_{\IL}^{t}$,
\begin{align*}
    \textbf{II}_{(13)} = & \frac{1}{n}\sum_{i=1}^n K_{h_l}(L_i -t)\big\{I\left(T_i \geqslant L_i\right)-g(\widebar{\Beta}_{\IL}^{t \trans} \Phibar_{\L i}^{t})\big\}\\
    & - \frac{1}{n}\sum_{i=1}^n K_{h_l}(L_i -t)\dot{g}(\widebar{\Beta}_{\IL}^{t\trans} \Phibar_{\L i}^{t})\Phibar_{\L i}^{t\trans} \left(\Beta_{\IL}^t-\widebar{\Beta}_{\IL}^t\right) + o_p\big(\|\widehat{\Beta}_{\IL}^t-\widebar{\Beta}_{\IL}^t\|_2 + (nh_l)^{-\frac{1}{2}}\big) = 0.
\end{align*}
Then we have 
\begin{align*}
    &\widehat{\Beta}_{\IL}^{t}= \arg\min\limits_{\Beta_{\L}^t}\left(\Beta_{\IL}^t-\widebar{\Beta}_{\IL}^t\right)^{\trans}\Big[-\textbf{A}_2\left(\Beta_{\IL}^t-\widebar{\Beta}_{\IL}^t\right)- \bm\Sigma_{21} +o_p\big(\|\widehat{\Beta}_{\IL}^t-\widebar{\Beta}_{\IL}^t\|_2 + (nh_l)^{-\frac{1}{2}}\big)\Big];\\
    &\qquad \qquad \text{s.t.}  \quad\textbf{B}_2^{\trans}\left(\Beta_{\IL}^t-\widebar{\Beta}_{\IL}^t\right) - \bm\Sigma_{22} - o_p\big(\|\widehat{\Beta}_{\IL}^t-\widebar{\Beta}_{\IL}^t\|_2 + (nh_l)^{-\frac{1}{2}}\big) =0.
\end{align*} 
where $\textbf{A}_2$ and $\textbf{B}_2$ are defined as above, and
\begin{align*}
    &\bm\Sigma_{21} =  \frac{h_l}{n}\sum_{i=1}^n \widebar{K}_{\NL}(t)^{-2} K_{h_l}^2(L_i -t)\big\{I\left(T_i \geqslant L_i\right)-g(\widebar{\Beta}_{\IL}^{t \trans} \Phibar_{\L i}^{t})\big\}\dot{g}(\widebar{\Beta}_{\IL}^{t\trans} \Phibar_{\L i}^{t})\Phibar_{\L i}^{t};\\
    &\bm\Sigma_{22}=\frac{1}{n}\sum_{i=1}^n K_{h_l}(L_i -t)\big\{I\left(T_i \geqslant L_i\right)-g(\widebar{\Beta}_{\IL}^{t \trans} \Phibar_{\L i}^{t})\big\}.
\end{align*}
Then from Theorem 5.21 of \citet{van2000asymptotic}, we obtain
\begin{align*}
    \widehat{\Beta}_{\IL}^t-\widebar{\Beta}_{\IL}^t
    = &\big[-\textbf{A}_2^{-1} + \textbf{A}_2^{-1}\textbf{B}_2(\textbf{B}_2^{\trans}\textbf{A}_2^{-1}\textbf{B}_2)^{-1}\textbf{B}_2^{\trans}\textbf{A}_2^{-1} \big]\bm\Sigma_{21} +\textbf{A}_2^{-1}\textbf{B}_2(\textbf{B}_2^{\trans}\textbf{A}_2^{-1}\textbf{B}_2)^{-1}\bm\Sigma_{22} \\
    = & O_p\big((nh_l)^{-\frac{1}{2}}\big).
\end{align*}
It then follows from the classical CLT that $\sqrt{nh_l}\> \big( \widehat{\Beta}_{\IL}^t-\widebar{\Beta}_{\IL}^t \big)$ is asymptotically normal. Hence we can use the same argument of the proof for the semi-supervised estimator $\Shat_{\L}(t)$ to show that $\sup_{t\in\mathcal{T}}\left|\Shat_\ssl-S(t)\right|=o_p(1)$. From \eqref{Lintrexp} in the main paper, we have
\begin{align}
\label{lintrasy}
&\sqrt{nh_l}\left( \Shat_\ssl - S(t) \right) \\\nonumber
=&\sqrt{nh_l} \Big[\frac{1}{N}\sum_{i=n+1}^{n+N}\widebar{K}_{\NL}(t)^{-1}K_{h_{\L}}(L_i -t)g(\widehat{\Beta}_{\IL}^{t\trans}\Phibar_{\L i}^{t}) - \frac{E\big[K_{h_{\L}}(L-t)g(\widebar{\Beta}_{\IL}^{t\trans}\Phibar_{\L}^{t})\big]}{E\left[K_{h_{\L}}(L-t) \right]} \\\nonumber
&+\frac{E\big[K_{h_{\L}}(L-t)g(\widebar{\Beta}_{\IL}^{t\trans}\Phibar_{\L}^{t})\big]}{E\left[K_{h_{\L}}(L-t) \right]} - S(t)\Big] \\\nonumber
= & \sqrt{nh_l}\Big[\frac{1}{N}\sum_{i=n+1}^{n+N}\widebar{K}_{\NL}(t)^{-1}K_{h_{\L}}(L_i -t)g(\widebar{\Beta}_{\IL}^{t\trans}\Phibar_{\L i}^{t})-\frac{E\big[K_{h_{\L}}(L-t)g(\widebar{\Beta}_{\IL}^{t\trans}\Phibar_{\L}^{t})\big]}{E\left[K_{h_{\L}}(L-t) \right]}\\\nonumber
&+\frac{1}{N}\sum_{i=n+1}^{n+N}\widebar{K}_{\NL}(t)^{-1}K_{h_{\L}}(L_i -t)\dot{g}(\widebar{\Beta}_{\IL}^{t\trans}\Phibar_{\L i}^{t}) \Phibar_{\L i}^{t\trans}\big(\widehat{\Beta}_{\IL}^{t}-\widebar{\Beta}_{\IL}^{t}\big) +o_p\big(\|\widehat{\Beta}_{\IL}^{t}-\widebar{\Beta}_{\IL}^{t}\|_2\big) \Big]\\\nonumber
=& \frac{1}{N}\sum_{i=n+1}^{n+N}f_l(t )^{-1}K_{h_{\L}}(L_i -t)\dot{g}(\widebar{\Beta}_{\IL}^{t\trans}\Phibar_{\L i}^{t}) \Phibar_{\L i}^{t\trans}\sqrt{nh_l}\big(\widehat{\Beta}_{\IL}^{t}-\widebar{\Beta}_{\IL}^{t}\big) + O_p\big(\sqrt{nh_l}/\sqrt{Nh_{\L}}\big)+o_p(1)\\\nonumber
=&f_l(t)^{-1} \sqrt{nh_l}\,\textbf{B}_{2}^{\trans}\big(\widehat{\Beta}_{\IL}^{t}-\widebar{\Beta}_{\IL}^{t}\big) + o_p(1)\\\nonumber
=&f_l(t)^{-1} \sqrt{nh_l}\,\bm{\Sigma}_{22} + o_p(1)\\\nonumber
=& \sqrt{\frac{h_l}{n}} \sum_{i=1}^n f_l(t)^{-1} K_{h_{\L}}(L_i -t)\big\{I\left(T_i \geqslant L_i\right)-g(\widebar{\Beta}_{\IL}^{t \trans} \Phibar_{\L i}^{t})\big\} +o_p(1).
\end{align}
It then follows from the classical CLT that
$$ \sqrt{nh_l}\left(\Shat_\ssl-S(t)\right)\stackrel{d}{\longrightarrow} N\left(0, \Sigma_\ssl\right), $$
where 
$\Sigma_\ssl= \nu_2 f_l(t)^{-1} E\big[I(T \geqslant t)-g(\widebar{\Beta}_{\IL}^{t \trans}\Phibar_{\L}^{t})\big]^2,$
which has a uniformly consistent estimator
\begin{equation*}
    \Sighat_\ssl=\frac{h_{l}}{n}\sum_{i=1}^{n} \widebar{K}_{\NL}(t)^{-2} K_{h_l}^2(L_i-t)\big\{I(T_i\geqslant L_i) - g(\bm{\widehat{\beta}}_{\IL}^{t \trans}\Phibar_{\L i}^{t})\big\}^2.
\end{equation*}
We then see that 
\begin{align*}
&\Sigma_\sl - \Sigma_\ssl = \nu_2 f_l(t)^{-1}E\big[I(T\geqslant t ) - S(t)\big]^{2} - \nu_2 f_l(t)^{-1}E\big[I(T\geqslant t) - g(\widebar{\Beta}_{\IL}^{t\trans}\Phibar_{\L}^{t})\big]^2\\
&=\nu_2 f_l(t)^{-1}E\Big[\big\{g(\widebar{\Beta}_{\IL}^{t\trans}\Phibar_{\L}^{t}) -S(t)\big\}^2 + 2 \big\{I(T\geqslant t) - g(\widebar{\Beta}_{\IL}^{t\trans}\Phibar_{\L}^{t})\big\}\big\{g(\widebar{\Beta}_{\IL}^{t\trans}\Phibar_{\L}^{t}) -S(t)\big\}\Big].
\end{align*}
Therefore, when the imputation model \eqref{lmodel} given in the main paper is correctly specified, it follows that
$$\Sigma_\sl - \Sigma_\ssl =\nu_2 f_l(t)^{-1}E\big[g(\widebar{\Beta}_{\IL}^{t\trans}\Phibar_{\L}^{t}) -S(t)\big]^2 \succeq \bm{0}.$$

\bibliographystyle{apalike}
\bibliography{reference}

\end{document}